\documentclass[12pt]{article}

\usepackage{enumerate}
\usepackage{natbib}
\usepackage{url} 
\usepackage{amsfonts,amsmath,amssymb,epsfig,epsf,psfrag}
\usepackage{graphicx}
\usepackage{xcolor}
\usepackage{xspace}

\usepackage[latin1]{inputenc}

\newcommand{\blind}{1}

\newcommand{\Yij}{Y_{ij}}
\newcommand{\xij}{x_{ij}}

\newcommand{\Bcal}{\mathcal{B}}

\newcommand{\dd}{\mathrm{d}}

\newcommand{\Esp}{\mathbb{E}}
\newcommand{\Dir}{\mbox{Dir}}

\newcommand{\Ncal}{\mathcal{N}}

\newcommand{\LowerBound}{{\mathcal L}}

\newcommand{\Trace}{\mathrm{Tr}}

\newtheorem{theorem}{Theorem}

\newtheorem{proposition}[theorem]{Proposition}

\begin{document}

\def\spacingset#1{\renewcommand{\baselinestretch}%
{#1}\small\normalsize} \spacingset{1.5}


%

\if1\blind
{
  \title{\bf Goodness of fit of logistic regression models for random graphs}
  \author{Pierre Latouche\hspace{.2cm}\\
    Laboratoire    SAMM,   EA   4543,    Université   Paris   1
  Panthéon-Sorbonne,  France\\
    and \\
    Stéphane Robin\\
    AgroParisTech, UMR 518, MIA, Paris, France\\
    INRA, UMR 518, MIA, Paris, France \\
  and \\
    Sarah Ouadah\\
    AgroParisTech, UMR 518, MIA, Paris, France\\
    INRA, UMR 518, MIA, Paris, France \\
}
  \maketitle
} \fi

\if0\blind
{
  \bigskip
  \bigskip
  \bigskip
  \begin{center}
    {\LARGE\bf Goodness of fit of logistic regression models for random graphs}
\end{center}
  \medskip
} \fi

\bigskip
\begin{abstract}

Logistic regression is a natural and simple tool to understand how covariates contribute to explain the topology of a binary network. Once the model fitted, the practitioner is interested in the goodness-of-fit of the regression in order to check if the covariates are sufficient to explain the whole topology of the network and, if they are not, to analyze the residual structure.
To address this problem, we introduce a generic model that combines logistic regression with a network-oriented residual term. This residual term takes the form of the graphon function of a $W$-graph. Using a variational Bayes framework, we infer the residual graphon by averaging over a series of blockwise constant functions. This approach allows us to define a generic goodness-of-fit criterion, which corresponds to the posterior probability for the residual graphon to be constant.
Experiments on toy data  are carried out to assess the accuracy of the procedure. Several networks  from social sciences and ecology are studied to illustrate the proposed methodology.

\end{abstract}

\noindent%
\vfill

\newpage
\spacingset{1.45} 



\section{Introduction}

Networks are now used in many scientific fields \citep{articlesnijders1997, articlewatts1998, articlenowicki2001, articlehoff2002, articlehandcock2007, articlezanghi2008} from biology \citep{articlealbert2002, Newman2003GraphSurveySIAM, articlebarabasi2004, articlelacroix2006} to historical sciences \citep{villa2008mining, jernite2014} and geography \citep{ducruet2013}. Indeed, while being simple data structures, they are yet capable of describing complex interactions between entities of a system. A lot of effort has been put, especially in social sciences, in developing methods to characterize the heterogeneity of these networks using latent variables, covariates, or both \citep{articlehoff2002,articlehandcock2007,articlemariadassou2010,articlezanghi2010}.

%


In this paper, we are interested in the contribution of covariates to explain the topology of an observed network. To this aim, we  consider standard logistic  models which are a simple way to account for the possible effect of covariates, assuming edges to  be independent conditionally on the covariates. Our goal is to provide the practitioners with tools to check the fit of the model and/or to analyze the residual structure. This goes along with the characterization of some residual structure present in the network that is not explained by the covariates. Our approach consists in combining logistic regression with the graphon function of a $W$-graph. This additional term plays the role of a very flexible, network-oriented residual term that can be visualized and on which a goodness-of-fit criterion can be based.

The $W$-graph can be casted among the latent-variable network models \citep{articlegoldenberg2009,matias2014}. It is characterized by a
function $W$ called \emph{graphon} where  $W(u, v)$ is the probability
for two nodes, with latent coordinates $u$ and $v$, each sampled from an
uniform   distribution  over   $[0,1]$,  to   connect.  As   shown  in
\cite{LoS06}, it is the limiting adjacency matrix of the network. This
result  comes  from  graph  limit theory  for  which 
\cite{DiJ08}  gave a  proper definition  using Aldous-Hoover theorem,
which  is   an  extension  of  de Finetti's   theorem  to  exchangeable
arrays. Until recently, few inference  techniques had been proposed to
infer the  graphon function  of a network.  The earliest  reference is \cite{kallenberg99}.   Since    then,   both   parametric
\citep{hoff2008,  PLV10}  and  non   parametric  \citep{Cha12} techniques  have  been
developed.  Graphon  inference is  a  particularly  challenging
problem  which has  received strong  attention is  the last  few 
years \citep{Cha12,ACC13,WoO13,asta14,ChA14,YHA14}.  In the present paper,
we follow \cite{LaR13} who took advantage of the fact that the well-known stochastic block model \citep[SBM:][]{articleholland1983,wang1987,articlenowicki2001} is a special case of $W$-graph corresponding to a blockwise constant graphon. This enables them to derive a variational Bayes EM (VBEM)  procedure to estimate the graphon function  as an  average of  SBM  models with  increasing
number of blocks. \bigskip

\noindent
As mentioned above, the model we consider combines a logistic regression term with a residual graphon function. 
Following the Bayesian framework of \cite{LaR13}, we estimate the residual graphon by averaging over a series of SBM including the one-block SBM, which corresponds to a constant residual graphon. We interpret a constant residual graphon as an absence of residual structure in the network. This approaches enables us 
\begin{enumerate}[($a$)]
 \item to assess the goodness-of-fit of the logistic regression through the posterior probability for the residual graphon to be constant and
 \item to display an estimate of the residual graphon that allows a visual inspection of the residual structure.
\end{enumerate}
As the exact Bayesian inference of this new  model  for networks  is not  tractable, we make an intensive use of variational Bayes approximations to achieve the inference. Because of the combination of logistic regression and SBM, two different types of variational approximations are actually required. \\

In Section \ref{sec:goodfit}, we introduce the general model and we define the goodness-of-fit criterion. Technical  issues and  theoretical aspects are  addressed in Section \ref{sec:inference}.  Finally, toy  and real  data sets  are analyzed  in Section \ref{sec:simu} and \ref{sec:examples} respectively to illustrate the proposed methodology.
In  the  body  of  the article,  only  undirected  networks  are considered.  The extension  to  directed networks  (with proofs  and update formulas) is derived in the supplementary materials. 
The proposed methodology is implemented in the R package GOFNetwork (github.com/platouche/gofNetwork),
which will be available on the Comprehensive R Archive Network (CRAN).

\section{Assessing goodness-of-fit}\label{sec:goodfit}

We consider a set of $n$ individuals among which interactions are observed. The observed interaction network is encoded in  the binary adjacency  matrix $Y = (\Yij)_{1\leq i, j \leq n}$ where $\Yij$ is 1 if nodes $i$ and $j$ are connected, and 0 otherwise. We further assume that a $d$-dimensional vector, $d\geq 1$, of covariates $\xij$ is available for each pair of nodes. In the following, we denote as $X = (\xij)_{1 \leq i, j \leq n}$ the set of all covariates. 

\subsection{Logistic regression and residual structure}\label{ssec:goodfit}
The influence of the covariates on  the network topology can be easily
accounted for using a logistic  regression model. Such a model assumes
that the edges $(\Yij)$ are independent (conditionally on the covariates) with respective distribution
\begin{equation*} \label{Eq:H0}
  H_0: \qquad Y_{ij} \sim \Bcal\left[g(x_{ij}^\intercal \beta + \alpha)\right],
\end{equation*}
where $\beta\in\mathbb{R}^d$, $\alpha\in\mathbb{R}$, $g$ stands for the
logistic  function $g(t)  = 1/(1  + \exp(-t))$,  $t\in\mathbb{R}$. Our
goal is to check if model $H_0$ is sufficient to explain the whole topology of the network. Note that the network structure does not explicitly appear in this model, as edges are considered as independent outcomes of a (generalized) linear model.

\bigskip
To assess the fit of Model $H_0$, we define a generic alternative network model. The alternative we consider is inspired from the $W$-graph model. More precisely, we consider the model
\begin{equation*} \label{Eq:H1}
  H_1: \qquad Y_{ij} \sim \Bcal\left[g(x_{ij}^\intercal \beta + \phi(U_i, U_j))\right],
\end{equation*}
where the $(U_i)_{1\leq i\leq n}$ are independent unobserved latent variables, with uniform distribution over the $(0, 1)$ interval. The non-constant function $\phi: (0, 1)^2 \mapsto \mathbb{R}$ encodes a residual structure in the network, that is not accounted for by Model $H_0$. Note that, in absence of covariate, this model corresponds to a $W$-graph \citep{LoS06} with graphon function $g \circ \phi$. Model $H_0$ corresponds to the case where the residual function $\phi$ is constant.

The present paper focuses on the goodness-of-fit of a regression model, therefore, the interpretation of the residual term $\phi(U_i, U_j)$ is not critical but its visual inspection may help to better understand where the residual heterogeneity does come from. {Note this generic form} encompasses additive node effect, which, in absence of regression term, would result in a model close to 
the expected degree model \citep{ChL02}.

\bigskip
The inference  of the function  $\phi$ in Model  $H_1$ is not  an easy
task and, following \cite{ACC13} and \cite{LaR13}, we consider a class of blockwise constant $\phi$ function. More precisely, we define the Model
\begin{equation} \label{Eq:MK}
  M_K: \qquad Y_{ij} \sim \Bcal\left[g(x_{ij}^\intercal \beta + Z_i^\intercal \alpha Z_j)\right],
\end{equation}
where $\alpha$ is a $K \times K$ real matrix ($K\geq 1$) and where the
$(Z_i)_{1  \leq   i  \leq  n}$   are  independent  vectors   with  $K$
coordinates, all  zero except one. We  denote $\pi_k$ ($1 \leq  k \leq
K$) the  probability that  the $k$th  coordinate is  non-zero. Briefly
speaking,   each    vector   $Z_i$   has    multinomial   distribution
$\mathcal{M}(1, \pi)$ where $\pi = (\pi_k)_{1 \leq k \leq K}$. The set
of parameters of such a model is $\theta = (\beta, \pi, \alpha)$. Note
that in the  absence of covariate, this model corresponds  exactly to a
SBM model. 
The ability of the stochastic block model to approximate the $W$-graph model is demonstrated in \cite{ACC13} and \cite{LaR13} and is not the purpose of this article.

Model $H_0$ is then equivalent to Model $M_1$ so the goodness-of-fit problem can be rephrased as the comparison between Model $H_0$ and $H'_1$, where
$$
H_0 = M_1 \qquad \text{and} \qquad H'_1 = \bigcup_{K \geq 2} M_K.
$$

\subsection{Bayesian model comparison}\label{ssection:bayesComp}

Now, we are given a series of Models $M_K$ ($K \geq 1$) indexed by $K$ which characterize $H_0$ and $H_1'$.
In this paper, we propose to compare $H_0$ and $H_1'$ using a Bayesian model comparison framework.

 Thus, each Model $M_K$ is associated to a prior probability $p(M_K)$. The parameter $\theta$ is then drawn conditionally on $M_K$ according to the prior distribution $p(\theta | M_K)$. Given $\theta$, $M_K$ and the given set $X$ of covariates, the graph is finally assumed to be sampled according to Model \eqref{Eq:MK}.
In this framework the prior probability of Models $H_0$ and $H'_1$ are 
$$
p(H_0) = p(M_1) \qquad \text{and} \qquad p(H'_1) = \sum_{K \geq 2} p(M_K).$$
Moreover, the posterior probability of Model $M_K$ is
\begin{equation}\label{eq:gofpmk}
  p(M_K |Y) = \frac{p(Y|M_K)p(M_K)}{p(Y)} = \frac{p(Y|M_K)p(M_K)}{\sum_{K'\geq 1}p(Y|M_{K'})p(M_{K'})}.
\end{equation}

\bigskip
The goodness of fit of Model $H_0$ can then be assessed by computing the posterior probability of $H_0$:
\begin{equation}
  \label{eq:postH0}
p(H_0|Y) = p(M_1|Y).  
\end{equation}
The Bayes factor \citep{KaR95} between Models $H_0$ and $H'_1$ can be computed in a similar way as
\begin{equation}
  \label{eq:BF}
B_{01} = \frac{p(Y|H_0)}{p(Y|H'_1)} 
\qquad \text{where} \quad
p(Y|H'_1) = \frac1{p(H'_1)} \sum_{K \geq 2} p(M_K) p(Y|M_K)  .  
\end{equation}

\section{Inference}\label{sec:inference}

The goodness-of-fit criteria introduced in the previous section all
depend on marginal likelihood terms $p(Y|M_K)$ which have to be estimated from
the data in practice. This is the object of this section. The prior distributions $p(M_K)$ and $p(\theta|M_K)$ are first
introduced. A variational three steps optimization scheme, based on global and
local variational methods, is then derived for inference.

In the following, we focus on undirected networks and therefore both
the adjacency matrix $Y$ and the matrix $X$ of covariates are
symmetric: $Y_{ij}=Y_{ji}$ and $x_{ij}=x_{ji},\forall i \neq
j$. Moreover, we
do not consider self-loops, \emph{i.e.} the connection of a node to
itself and therefore the pairs $(i, i),\forall i$ are discarded from the sums
and products involved. The complete derivation of the model and the inference procedure
in  the  directed  case  are given  as  supplementary  materials.  The
Appendix with  all proofs in the  undirected case is also  provided as
supplementary materials.

\subsection{Prior distributions}

With no prior information on which model should be preferred, we give
equal weights $p(H_0)=p(H_1')=1/2$ to $H_0$ and $H_1'$. Therefore,
$p(M_1)=1/2$. Alternative
choices can be made by integrating expert knowledge at hand. Recall
that $p(H_1')=\sum_{K \geq 2}p(M_K)$.

For Model $M_K$, the prior distribution over the model parameters in
$\theta$ is defined as a product of conjugate prior
distributions over the different sets of parameters: $p(\theta
|M_K)=p(\beta|M_K)p(\pi|M_K)p(\alpha|M_K)$. Since $\pi$ is involved in
a multinomial distribution to sample the vectors $Z_i$, a Dirichlet
prior distribution is chosen
\begin{equation*}
 p(\pi |M_K) = \mathrm{Dir}(\pi; e),
\end{equation*}
where $e$ is a vector with $K$ components such that
$e_{k}=e_{0}>0,\forall k \in \{1,\dots,K\}$. Note that fixing
$e_0=1/2$ induces a non-informative Jeffreys prior distribution which
is known to be proper \citep{proceedingsjeffreys1946}. It is also
possible to obtain a uniform distribution over the $K-1$ dimensional
simplex by setting $e_0=1$.

In order to characterize the $d$-dimensional regression vector
$\beta$, a Gaussian distribution is considered
\begin{equation*}
 p(\beta|\eta,M_K)
 = \Ncal\left(\beta;\:0,\frac{I_{d}}{\eta}\right)
 = \prod_{j=1}^{d}\Ncal\left(\beta_j;\:0,\frac{1}{\eta}\right),
\end{equation*}
with $I_{d}$ the $d \times d$ identity matrix and $\eta>0$ a parameter
controlling the inverse variance. Similarly, the matrix $\alpha$ is
modeled using a product of Gaussian distributions with $\gamma>0$
controlling the variance
\begin{equation*}
 p(\alpha|\gamma,M_K)=\prod_{k \leq l}^K \Ncal\left(\alpha_{kl};\:0,\frac{1}{\gamma}\right).
\end{equation*}
Since we focus on undirected networks, $\alpha$ has to
be symmetric and therefore the product involves the $k \leq l$ terms
of $\alpha$.
In the directed case (see supplementary materials), the product is
over all terms $k, l$ and the $\mathrm{vec}$ operator, which stacks the
columns of a matrix into a vector, is used to simplify the
calculations. 

Finally, Gamma distributions are considered for $\gamma$
\begin{equation*}
 p(\gamma|M_{K})=\mathrm{Gam}(\gamma;\:a_{0},b_{0}), \quad a_{0},b_{0}>0,
\end{equation*}
and $\eta$
\begin{equation*}
 p(\eta|M_{K})=\mathrm{Gam}(\eta;\:c_{0},d_{0}), \quad c_{0},d_{0}>0.
\end{equation*}
By construction, Gamma distributions are informative. In order to
limit the influence on the posterior distributions, the
hyperparameters controlling the scale ($a_0, c_0$) and rate ($b_0,
d_0$) are usually set to low values in the literature.

The choice of modeling the prior information on the
parameters 
$\alpha$ and $\beta$ from such Gaussian-Gamma distributions 
 has been widely considered both in standard Bayesian linear
regression and Bayesian logistic regression 
\citep[see for instance][]{proceedingsbishop2003,bookbishop2006}. The
prior distributions $p(\beta|M_K)$ and $p(\alpha|M_K)$ are then
obtained by marginalizing over $p(\eta|M_K)$ and $p(\gamma |M_K)$
respectively. This results in prior distributions from the class of generalized hyperbolic
distributions. For more details, we refer to \cite{caron2008}.

In the following, and in order to simplify the notations, the
dependency on $M_K$ is omitted in the prior and posterior
distributions. 

\subsection{Variational approximations}

Denoting $Z$ the set of all latent vectors $(Z_i)$, the
marginal log-likelihood of Model $M_K$, also called the
integrated observed data log-likelihood, is given by
\begin{equation}\label{eq:margin}
\log p(Y|M_K) = \log\left\{\sum_{Z}\int
 p(Y|Z,\alpha,\beta)p(Z|\pi)p(\alpha|\gamma)p(\beta|\eta)p(\pi)p(\gamma)p(\eta)\dd\pi
\dd\alpha \dd\beta \dd\gamma \dd\eta\right\}.
\end{equation}
It requires a marginalization over the prior distributions of all
parameters. In particular, it involves testing all the $K^n$
configurations of $Z$.
Unfortunately, (\ref{eq:margin}) is not tractable and therefore we
propose to rely on variational approximations for inference purposes. Let us first consider the
global variational decomposition
\begin{equation}\label{eq:decomp1}
\log p(Y|M_K)= \LowerBound_K(q) + \mathrm{KL}\left(q(\cdot)||p(\cdot|Y, M_K)\right).
\end{equation}
Maximizing the functional $\LowerBound_K(\cdot)$, which is a lower bound
of $\log p(Y|M_K)$, with respect to the distribution $q(\cdot)$, is
equivalent to minimizing the Kullback-Leibler divergence between
$q(\cdot)$ and the unknown posterior distribution
$p(\cdot|Y)$. $\LowerBound_K(\cdot)$ is given by
\begin{equation*}
\LowerBound_K(q)=\sum_{Z}\int
q(Z,\pi,\alpha,\beta,\gamma,\eta)\log \frac{p(Y,Z,\pi,\alpha,\beta,\gamma,\eta)}{q(Z,\pi,\alpha,\beta,\gamma,\eta)}\dd\pi
\dd\alpha \dd\beta \dd\gamma \dd\eta.
\end{equation*}
In order to maximize the lower bound, we assume that the distribution
can be factorized as follows:
\begin{equation*}
q(Z,\pi,\alpha,\beta,\gamma,\eta)=q(\pi)q(\alpha)q(\beta)q(\gamma)q(\eta)\prod_{i=1}^{n}q(Z_{i}).
\end{equation*}
Unfortunately, $\LowerBound_K(\cdot)$ is still intractable due to the
logistic function in $p(Y|Z,\alpha, \beta)$. Following the work of
\cite{articlejaakkola2000}, a tractable lower bound is derived.

\begin{proposition}\label{prop:bound}
 Given any $n \times n$ positive real matrix
 $\xi=(\xi_{ij})_{1\leq i,j\leq n}$, a lower bound of the first lower bound is given by
\begin{equation*}
\log p(Y|M_K) \geq \LowerBound_K(q) \geq \LowerBound_K(q;\:\xi),
\end{equation*}
where 
\begin{equation*}
\LowerBound_K(q;\:\xi)=\sum_{Z}\int
q(Z,\pi,\alpha,\beta,\gamma,\eta)\log \frac{\sqrt{h(Z,\alpha,\beta,\xi)}p(Z,\pi,\alpha,\beta,\gamma,\eta)}{q(Z,\pi,\alpha,\beta,\gamma,\eta)}\dd\pi
\dd\alpha \dd\beta \dd\gamma \dd\eta,
\end{equation*}
and 
\begin{multline*}
\log h(Z,\alpha,\beta,\xi) = \sum_{i \neq
 j}^{n}\Big\{(Y_{ij}-\frac{1}{2})(Z_{i}^{\intercal}\alpha Z_{j}+x_{ij}^{\intercal}\beta)
 + \log g(\xi_{ij}) - \frac{\xi_{ij}}{2} \\ -
 \lambda(\xi_{ij})\left((Z_{i}^{\intercal}\alpha Z_{j} +
 x_{ij}^{\intercal}\beta)^{2} - \xi_{ij}^{2}\right)\Big\},
\end{multline*}
with $\xi_{ij}\in \mathbb{R}^{+}$, $\xi_{ij} = \xi_{ji}$. Moreover, 
$\lambda(\xi_{ij})=\left(g(\xi_{ij})-1/2\right)/(2\xi_{ij})$, $g$ being the logistic function. 
\end{proposition}
The proof is given in Appendix A.1.
The quality of the lower bound $\LowerBound_K(q;\:\xi)$, which was
obtained through a series of Taylor expansions, clearly depends on the
choice of the matrix $\xi$. As we shall see in Section \ref{ssection:optimxi}, $\xi$ can
be estimated from the data to obtain tight bounds.

\subsubsection{Variational Bayes EM}

For now, we assume that the matrix $\xi$ is fixed and we rely on
$\LowerBound_K(q;\:\xi)$ as a lower bound of $\log p(Y|M_K)$. In order to
maximize the lower bound, a VBEM algorithm \citep{proceedingsbeal03} is applied on
$\LowerBound_K(q;\:\xi)$. This optimization scheme is iterative and is
related to the EM algorithm \citep{articledempster1977}. Keeping all
distributions fixed except one, the bound is maximized with respect to
the remaining distribution. This procedure is repeated in turn until
convergence of the bound. The optimization of the distribution $q(Z)$ over
the latent variables usually refers to the variational E step. The
updates of $q(\pi)$, $q(\alpha)$, $q(\beta)$, $q(\gamma)$, 
and  $q(\eta)$  refer here  to  the  variational M  step.  Proposition
\ref{prop:Zi}  provides   the  update   formula  of  the   E-step  and
Propositions \ref{prop:pi}  to \ref{prop:alpha}  provide these  of the
M-step. The corresponding proofs are given in
Appendix A.2 to A.7.

\begin{proposition}\label{prop:Zi}
 The variational E update step for each distribution $q(Z_i)$ is
 given by:
\begin{equation*}
q(Z_i) = \mathcal{M}(Z_i;\:1,\tau_{i}),
\end{equation*}
where $\sum_{k=1}^K \tau_{ik}=1$ and 
\begin{multline*}
\tau_{ik} \propto \exp \Bigg\{ 
 \sum_{l=1}^{K}(m_{\alpha})_{kl} \sum_{j \neq i}^n
 \Big((Y_{ij}-\frac{1}{2}) - 2
 \lambda(\xi_{ij})x_{ij}^{\intercal}m_\beta \Big)\tau_{jl} - \sum_{l=1}^K \Esp_{\alpha_{kl}}[\alpha_{kl}^2]\sum_{j \neq i}^n
 \lambda(\xi_{ij}) \tau_{jl} \\ 
+ \psi(e_k^n) -
 \psi\Big(\sum_{l=1}^K e_l^n\Big) \Bigg\}.
\end{multline*}
$\psi(\cdot)$ denotes the digamma function which is the logarithmic
derivative of the gamma function.
\end{proposition}

\begin{proposition} \label{prop:pi}
 The variational M update step for the distribution $q(\pi)$ is
 given by:
\begin{equation*}
q(\pi) = \Dir(\pi;\:e^{n}),
\end{equation*}
where, $\forall k \in \{1,\dots,K\}$, $e_{k}^{n}=e_{0} + \sum_{i=1}^{n}\tau_{ik}$, $\tau_{ik}$ being given by Proposition \ref{prop:Zi}. 
\end{proposition}

\begin{proposition}\label{prop:beta}
 The variational M update step for the distribution $q(\beta)$ is
 given by:
\begin{equation*}
q(\beta) = \mathcal{N}(\beta;\:m_\beta, S_\beta),
\end{equation*}
where
\begin{equation*}
S_\beta^{-1} = \frac{c_n}{d_n}I_{d} + \sum_{i \neq j}^n
\lambda(\xi_{ij}) x_{ij}x_{ij}^{\intercal},
\end{equation*}
and 
\begin{equation*}
m_{\beta} = S_{\beta}\frac{1}{2}\sum_{i \neq j}^n
\left(Y_{ij}-\frac{1}{2} - 2\lambda(\xi_{ij})\tau_i^\intercal
 m_\alpha \tau_j \right) x_{ij}.
\end{equation*}
\end{proposition}

\begin{proposition}\label{prop:gamma}
 The variational M update step for the distribution $q(\gamma)$ is
 given by:
\begin{equation*}
q(\gamma) = \mathrm{Gam}(\gamma;\:a_{n},b_{n}),
\end{equation*}
where $a_{n}=a_0 + \frac{K(K+1)}{4}$ and $b_n = b_0 + \frac{1}{2}\sum_{k\leq l}^{K} \Esp_{\alpha_{kl}}[\alpha_{kl}^{2}]$.
\end{proposition}

\begin{proposition}\label{prop:eta}
 The variational M update step for the distribution $q(\eta)$ is
 given by:
\begin{equation*}
q(\eta)=\mathrm{Gam}(\eta;\:c_{n},d_{n}),
\end{equation*}
where $c_{n}=c_0 + \frac{d}{2}$ and $d_n = d_0 +
\frac{1}{2}\Trace(S_\beta) + \frac{1}{2}m_\beta^\intercal
 m_\beta$, $S_\beta$ and $m_\beta$ being given by Proposition \ref{prop:beta}.
\end{proposition}

\begin{proposition}\label{prop:alpha}
 The variational M update step for the distribution $q(\alpha)$ is
 given by:
\begin{equation*}
 q(\alpha) = \prod_{k \neq l}^{K} \mathcal{N}\left(\alpha_{kl};(m_{\alpha})_{kl},(\sigma_{\alpha}^2)_{kl}\right),
\end{equation*} 
where
\begin{equation*}
 (\sigma_{\alpha}^{2})_{kk}^{-1} = \frac{a_{n}}{b_{n}} + \sum_{i \neq
 j}^{n} \lambda(\xi_{ij})\tau_{ik}\tau_{jk},\forall k,
\end{equation*}
\begin{equation*}
 (\sigma_{\alpha}^{2})_{kl}^{-1} = \frac{a_{n}}{b_{n}} + 2\sum_{i \neq
 j}^{n} \lambda(\xi_{ij})\tau_{ik}\tau_{jl},\forall k \neq l,
\end{equation*}
\begin{equation*}
 (m_{\alpha})_{kk} = (\sigma_{\alpha}^{2})_{kk}\sum_{i \neq j}^{n}
 \left(\frac{1}{2}(Y_{ij}-\frac{1}{2}) -
 \lambda(\xi_{ij})x_{ij}^{\intercal}m_\beta\right)\tau_{ik}\tau_{jk},\forall k,
\end{equation*}
\begin{equation*}
 (m_{\alpha})_{kl} = (\sigma_{\alpha}^{2})_{kl}\sum_{i \neq j}^{n}
 \left((Y_{ij}-\frac{1}{2}) -
 2\lambda(\xi_{ij})x_{ij}^{\intercal}m_\beta\right)\tau_{ik}\tau_{jl},
 \forall k \neq l.
\end{equation*}
\end{proposition}

\subsubsection{Optimization of $\xi$}\label{ssection:optimxi}

So far, we have seen how the lower bound $\LowerBound_K(q;\:\xi)$ of
$\log p(Y|M_K)$ could be maximized with respect to the distribution
$q(Z, \pi, \alpha, \beta, \gamma, \eta)$. However, we have not
addressed yet how $\xi$ could be estimated from the data. Given a
distribution $q(\cdot)$, we propose to maximize $\LowerBound_K(q;\:\xi)$
with respect to each variable $\xi_{ij}$ in order to obtain the
tightest bound $\LowerBound_K(q;\:\xi)$ of $\log p(Y|M_K)$. This
follows the work of \cite{proceedingsbishop2003} on Bayesian
hierarchical mixture of experts and
\cite{articlelatouche2011,latouche2014} on the overlapping stochastic
block model. As shown in the following proposition, this leads to
new estimates $\widehat{\xi}_{ij}$ of $\xi_{ij}$.
\begin{proposition}\label{prop:xi}
The estimate $\widehat{\xi}_{ij}$ of $\xi_{ij}$ maximizing
$\LowerBound_K(q;\:\xi)$ is given by
 \begin{equation*}
\xi_{ij} = \sqrt{\sum_{k,l}^{K}\tau_{ik}\tau_{jl}\Esp_{\alpha_{kl}}[\alpha_{kl}^{2}] + 2\sum_{k,l}^{K}\tau_{ik}\tau_{jl}(m_{\alpha})_{kl}x_{ij}^{\intercal}m_{\beta}\\
+
 \Trace(x_{ij}x_{ij}^{\intercal}(S_{\beta}+m_{\beta}m_{\beta}^{\intercal}))}.
\end{equation*}
Note that $\widehat{\xi}_{ij}=\widehat{\xi}_{ji},\forall i \neq j$ since the
networks considered are undirected. 
\end{proposition}

This gives rise to a three steps optimization scheme. Given a matrix
$\xi$, the variational E and M steps of the VBEM algorithm are used to maximize $\LowerBound_K(q;\:\xi)$
with respect to $q(\cdot)$. This distribution is then held fixed and
the bound is maximized with respect to $\xi$. These three steps are
repeated until convergence of the lower bound. The proof is given in
Appendix A.8.

\subsection{Estimation}

\paragraph{Goodness-of-fit}

For any $K$, we have seen how variational techniques could be used to
approximate the marginal log-likelihood $\log p(Y|M_K)$ using a lower
bound $\widehat{\LowerBound}_K := \max_{q, \xi} \LowerBound_K(q, \xi)$. 
As exposed in Section \ref{ssec:goodfit}, our goodness-of-fit procedure relies on the posterior probability of $K$, that is $p(M_K|Y)$. Indeed, this posterior distribution cannot be derived in a exact manner but, as shown in \cite{VMR12}, the distribution $\widehat{p}(M_K|Y)$ that minimizes the Kullback-Leibler divergence with $p(M_K|Y)$ satisfies 
$$
\widehat{p}(M_K|Y) \propto p(M_K) \exp\{\widehat{\LowerBound}_K\}.
$$
The approximate posterior probability of $H_0$ is then $\widehat{p}(H_0|Y)=\widehat{p}(M_1|Y)$ and the corresponding approximate posterior Bayes factor $\widehat{B}_{01}$, defined in \eqref{eq:BF}, can be computed in the same manner.
 
The following proposition, which is proved in Appendix A.9, shows
that many terms of $\LowerBound_K(q;\:\xi)$ vanish, when computed after
a specific optimization step, so that the lower bound takes a simpler form.

\begin{proposition}\label{prop:boundM}
If computed right after the variational M step, the lower bound is
given by
 \begin{multline*}
\LowerBound_K(q;\:\xi) = \frac{1}{2}\sum_{i \neq j}^n \left\{ \log g(\xi_{ij}) -
 \frac{\xi_{ij}}{2} + \lambda(\xi_{ij})\xi_{ij}^2\right\} + \log
\frac{C(e^n)}{C(e)} + \log \frac{\Gamma(a_n)}{\Gamma(a_0)}+ \log \frac{\Gamma(c_n)}{\Gamma(c_0)}\\
+ a_0\log b_0 + a_n (1 -
\frac{b_0}{b_n} - \log b_n) 
+ c_0 \log d_0 + c_n(1 - \frac{d_0}{d_n} - \log d_n) \\
+ \frac{1}{2}\sum_{k \leq l}^{K}\log (\sigma_\alpha^{2})_{kl} + \frac{1}{2}\log |S_{\beta}| -
\sum_{i=1}^n\sum_{k=1}^K \tau_{ik}\log \tau_{ik} +
\frac{1}{2}\sum_{k \leq l}^{K}(\sigma_{\alpha}^{2})_{kl}^{-1}(m_{\alpha})_{kl}^{2}
- \frac{1}{2}m_{\beta}^{\intercal}S_{\beta}^{-1}m_\beta \\ 
+ \frac{1}{2}m_{\beta}^{\intercal}\sum_{i \neq j}^n
(Y_{ij}-\frac{1}{2})x_{ij},
\end{multline*}
where $C(x)=\prod_{k=1}^{K}\Gamma(x_{k}) \left/ \Gamma\left(\sum_{k=1}^{K}x_{k}\right) \right.$
and $\Gamma(\cdot)$ is the gamma function.
\end{proposition}

\paragraph{Residual structures}

While the main object of this work is to provide tools to assess the
goodness of fit of a logistic regression model for networks, the {considered}
variational algorithm also provides a natural way to estimate the
residual structure $\phi$. We recall that, {under}
Model $H_0$, \emph{i.e.} the network is completely explained by the
covariates, the function $\phi$ is constant.

Still,  under  the  alternative  Model  $H_1$,  a  residual  structure
remains, that is encoded in $\phi$. As a consequence,
an estimate of this function can be useful to investigate the residual
structure,  similarly  to the  residual  plot  classically used  in  a
regression context.  Removing the  covariate effect, recall that $M_K$
is  a SBM  model.  Therefore,  an approximate  posterior  mean can  be
derived, relying  on the VBEM  model averaging approach  considered in
\cite{LaR13} for SBM. Proposition \ref{prop:ma1} provides the
approximate posterior mean of the  function $\phi$, that we propose as
the network counterpart of the residual plot in regression. Note that it results from an integration over all model
parameters and Models $M_{K}$.

\begin{proposition}\label{prop:ma1}
 From Proposition 1 in \cite{LaR13}, for $(u,v)\in [0,1]^{2}, u \leq v$, the approximate
 posterior mean of the residual structure $\phi$ is 
 \begin{equation*}
 \widehat{\Esp}\big[\phi(u, v)|Y\big]=\sum_{K \geq 1}\widehat{p}(M_K|Y)\widehat{\Esp}\big[\phi(u,v)|Y,M_K\big],
 \end{equation*}
where
\begin{equation*}
 \widehat{\Esp}\big[\phi(u,v)|Y,M_K\big]=\sum_{k \leq l}(m_{\alpha})_{kl}\left[F_{{k-1}, {l-1}}(u, v; e) - F_{{k},
 {l-1}}(u, v; e) - F_{{k-1}, {l}}(u, v; e) + F_{{k},
 {l}}(u, v; e)\right].
\end{equation*}
 $F_{k,l}(u,v;e)$ denotes the joint cdf of the Dirichlet variables
$(\sigma_{k},\sigma_{l})$ such that $\sigma_{k}=\sum_{l=1}^{k}\pi_{l}$
and $\pi$ has a Dirichlet distribution $\Dir(e)$. 
\end{proposition}
As  mentioned   in  Section  \ref{ssec:goodfit},   the  residual
  structure $\phi$  is related  to the  graphon function  of $W$-graph
  models, which suffer from identifiability issues.  Indeed, for any
  measure preserving transformation $\sigma$ of  $[0, 1]$ to $[0, 1]$,
  the   function    $\phi_{\sigma}(u,   v)    =   \phi\left(\sigma(u),
    \sigma(v)\right)$ leads  to the  same model  as with  the function
  $\phi(u, v)$. To  tackle this issue, the  common approach is
  to assume that  the mean function $\int \phi(u,  v) \dd v$ is increasing
  in  $u$.  This identifiability  constraint  was  applied
  when  producing  the  residual  structure  plots  presented  in  the
  following section.


\section{Simulation study}\label{sec:simu}

In order to assess the proposed methodology, we carried out a series of experiments on simulated data first and then on
real data. In this section, we focus on the estimation of the posterior probability
$\hat{p}(H_0|Y)$. We aim at evaluating the capacity of the approach to
detect $H_1$ using toy data.  Similar results were obtained for the
estimated Bayes
factors $\hat{B}_{01}$ and identical conclusions were drawn. 

\subsection{Simulation design}

We simulated networks using Model $H_1$. Thus, each node is first associated
to a latent position $U_{i}$  sampled from a uniform distribution over
the  $(0,1)$  interval.   Then,  a  vector  of   covariates  $x_i  \in
\mathbb{R}^{d}$ is drawn for each node, using a standardized Gaussian distribution,
\emph{i.e.} with zero  mean and covariance matrix set  to the identity
matrix, 
with  $d=2$. In order  to construct the  covariate vector
$x_{ij}\in \mathbb{R}^{d}$ for each edge  $(i,j)$ with $(i<j)$, we fixed
$x_{ij}=x_{i}-x_{j}$. For the function $\phi(\cdot,\cdot)$, we considered a design inspired by the one proposed in 
\cite{LaR13}.  In this work, the graphon function is $W(u,
v)=\rho\lambda^{2}(uv)^{\lambda-1}$   where   the   parameter   $\rho>0$
controls the graph density and $\lambda>0$ the degree concentration. For
more  details, we  refer to  \cite{LaR13}.  Note  that the  maximum of
the graphon function is $\rho\lambda^{2}$ so $\lambda <
1/\sqrt{\rho}$ must hold since $W(\cdot,\cdot)$ is a probability. In
our case, the  probabilities for nodes to connect are  given through a
logistic function $g(\cdot)$ and therefore we set $\phi(u, v)
=g^{-1}\big(\rho\lambda^{2}(uv)^{\lambda-1}\big)$.  For  $\lambda=1$,  the  function
$\phi(\cdot,\cdot)$  is  constant and  so  the  networks are  actually
sampled from Model $H_0$. Conversely, for all $\lambda > 1$, data sets
come from Model $H_1$. As $\lambda$ increases, the residual structure,
not accounted for by Model $H_0$, becomes sharper and thus easier to detect.

We considered  networks of size $n=100$  and $n=150$ as well  as three
values   for   the  parameter   $\rho\in\{10^{-2},10^{-1.5},10^{-1}\}$
helping controlling the sparsity. Finally, we tested 20 different values of
$\lambda$  in  $[1,5]$. For each of the  triplets ($n,\rho, \lambda$),
we simulated 100 networks and we applied the methodology we propose for
values of $K$ between $1$  and $10$. Because the variational algorithm
depends on the initialization, as any  EM like procedure, for each $K$
it was run  twice and the best run was selected, such that the lower
bound was maximized. Note that equal prior probabilities were given for the Models $M_K$
($K   \geq   2$)   such   that   $p(H_1')=1/2$.   Moreover,   we   set
$a_0=b_0=c_0=d_0=e_0=1$.

\subsection{Results}

\begin{figure}[!] 
\centering
\caption{\label{fig:simugoofit} 
Boxplots  of the  estimated values  $\hat{p}(H_0|Y)$ of  the posterior
probability $p(H_0|Y)$, obtained with the variational
approximations, for values of $\lambda$ ranging from $1$ to
$5$. Six  scenarios  considered with  the number  $n$ of  nodes in
$\{100,  150\}$  and  the  sparsity parameter  $\rho$  in  $\{10^{-2},
10^{-1.5}, 10^{-1}\}$.  Model $H_0$  is true for  $\lambda=1$ and
false for $\lambda > 1$.}
\begin{tabular}[htbp]{c c c}
& $n=100$ nodes & $n=150$ nodes \\
 \rotatebox{90}{\hspace{0.2cm } High sparsity $(\rho=10^{-2})$} & \includegraphics[width=5.5cm]{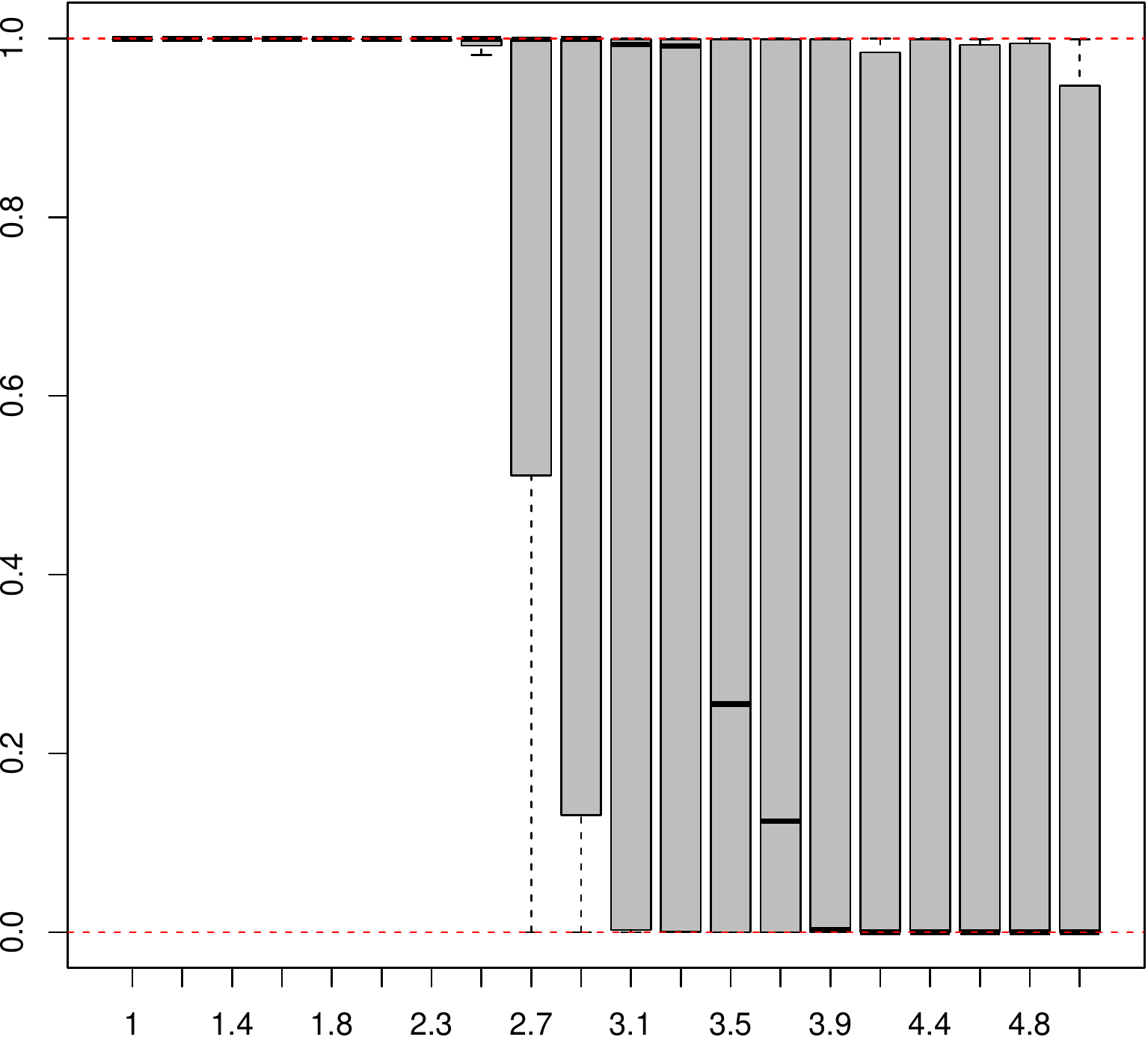}
       & \includegraphics[width=5.5cm]{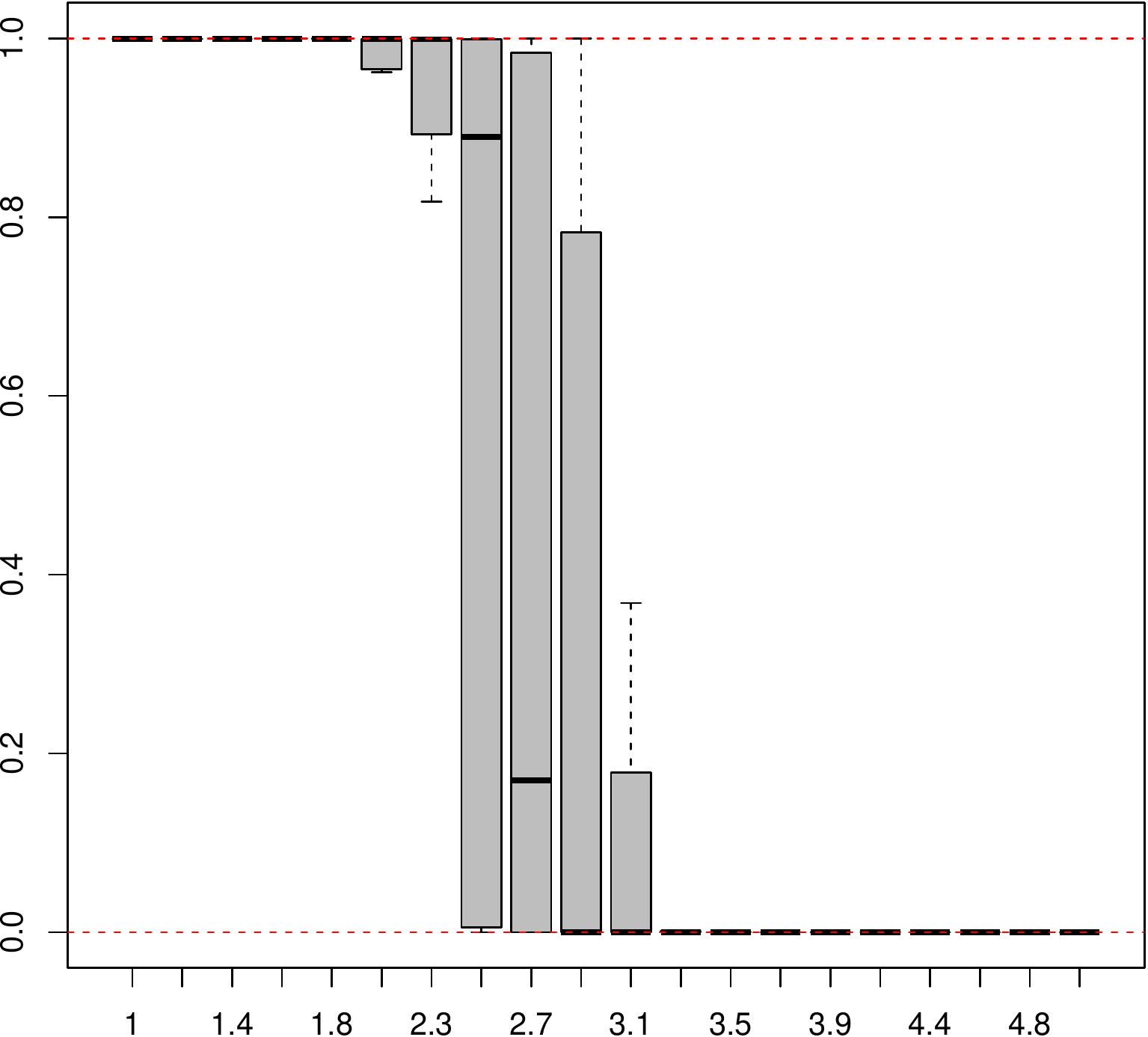} \\
\rotatebox{90}{\hspace{0.2cm } Average sparsity $(\rho=10^{-1.5})$} & \includegraphics[width=5.5cm]{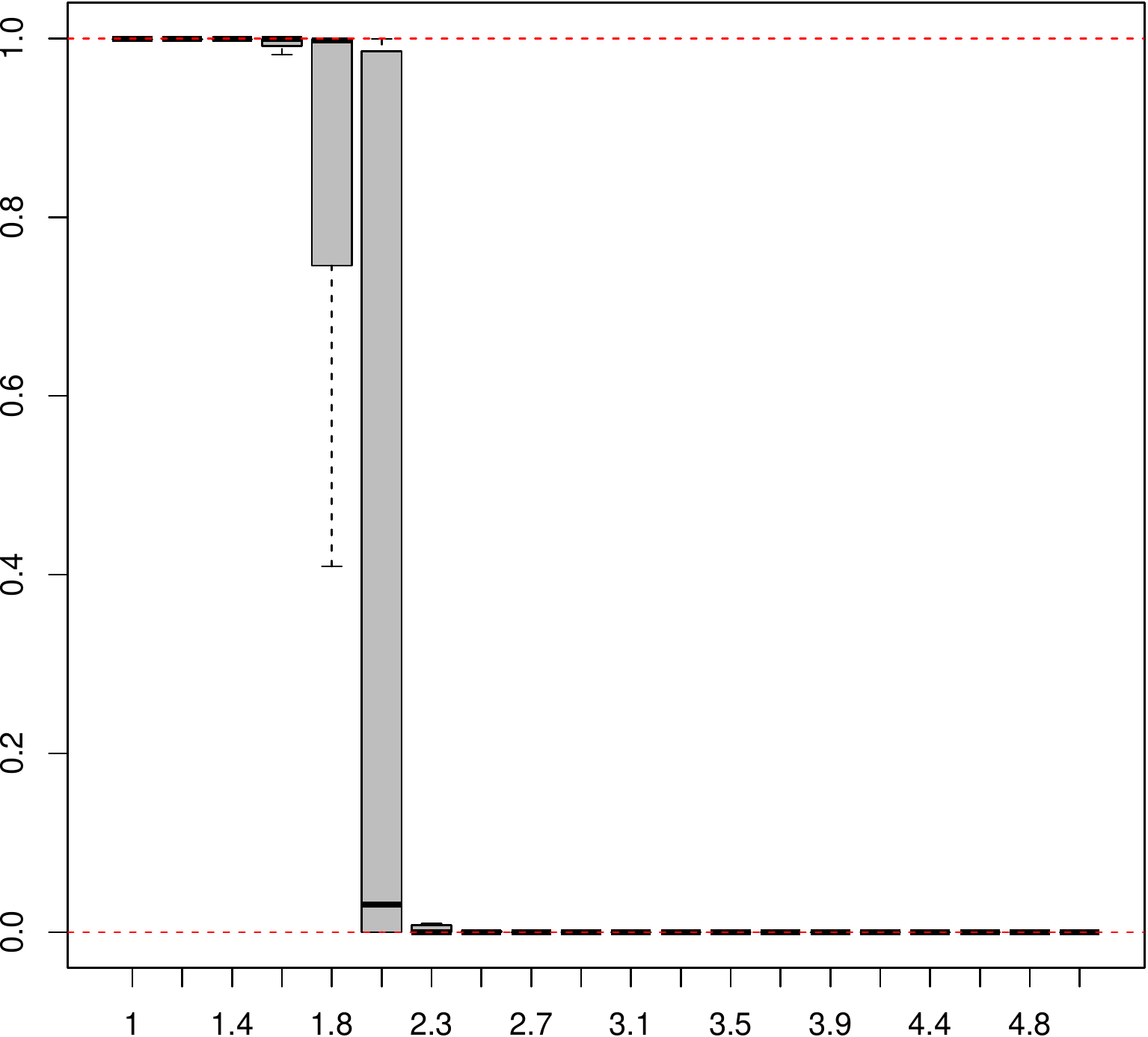}
       & \includegraphics[width=5.5cm]{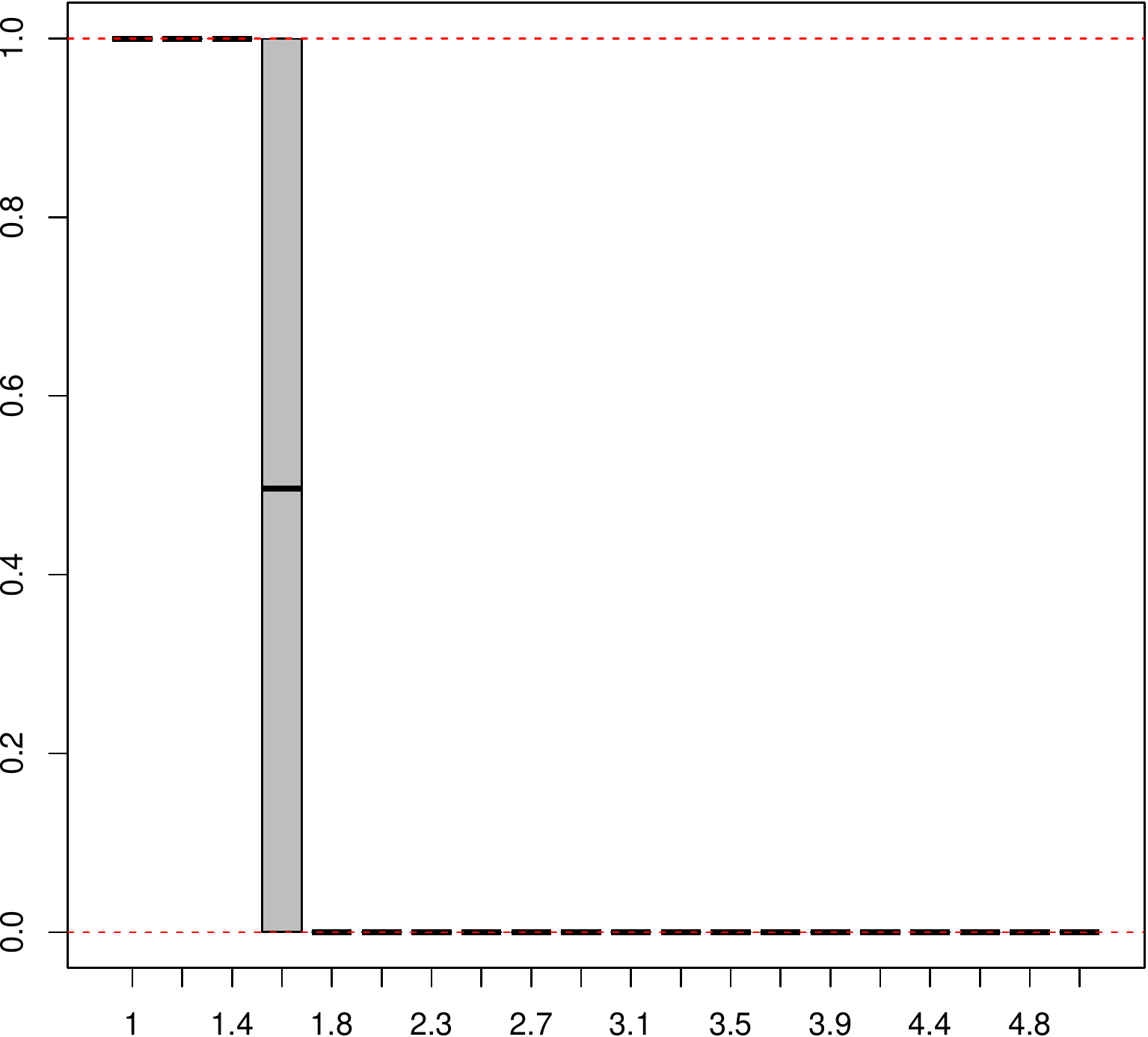}  \\
\rotatebox{90}{\hspace{0.2cm } Low sparsity $(\rho=10^{-1})$} & \includegraphics[width=5.5cm]{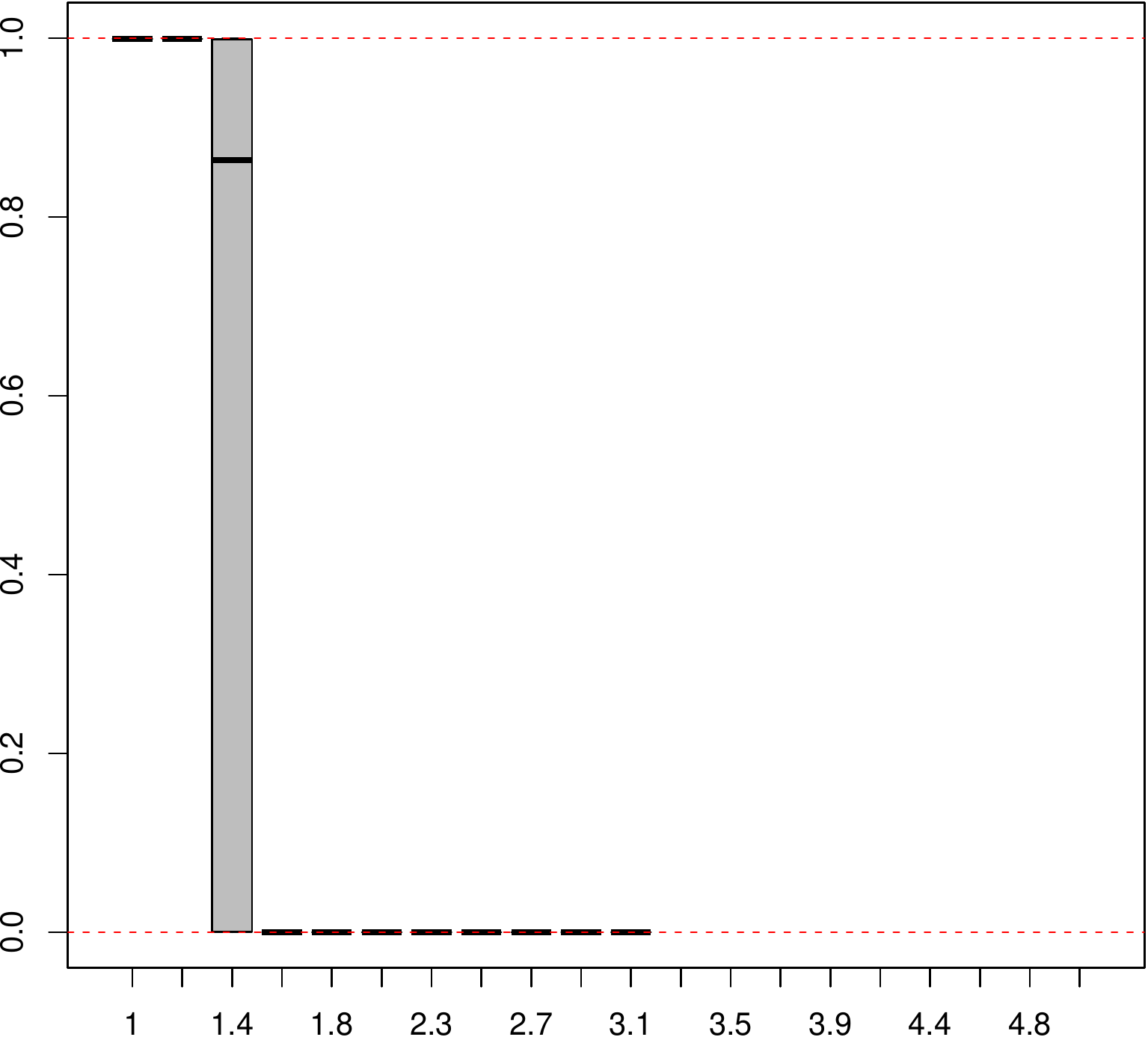} & \includegraphics[width=5.5cm]{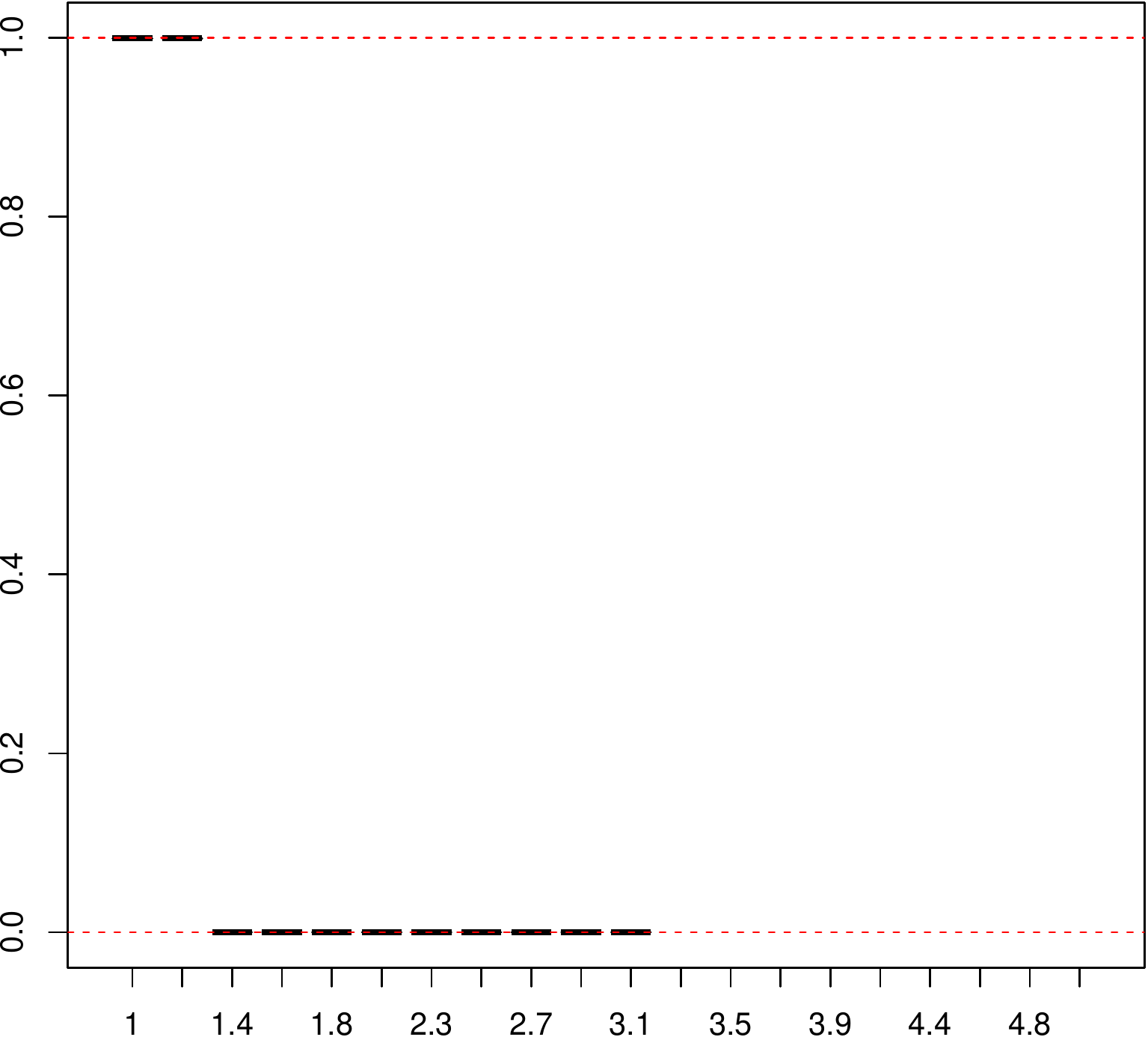} 
\end{tabular}
\end{figure}

\paragraph{Estimation of $p(H_0|Y)$.}
The   results  are   presented  in   Figure  \ref{fig:simugoofit}. It appears that  for low values of  $\lambda$, the
median (indicated in bold on the boxplots) of the 
estimated values of  $p(H_0|Y)$ is  1 and  goes to 0, when
$\lambda$ increases, as expected.  The results for the scenario with the highest
sparsity  ($\rho=10^{-2}$)  and   $n=100$  are   unstable
although  the median  values share this global  property. Much
stable  results  were  obtained  for  larger  networks.  Interestingly,
experiments  can   be  distinguished  in   the  way  Model   $H_1$  is
detected. As soon  as $\lambda>1$, then the true  model responsible for
generating the  data is $H_1$  and so  the probability of  Model $H_0$
should be lower than $1/2$. In practice, the estimated probability
$\hat{p}(H_0|Y)$ is lower than $1/2$ for slightly larger values of $\lambda$. For
instance, for
$\rho=10^{-1.5}$   and  $n=150$,   $\hat{p}(H_0|Y)\approx  0$   for
$\lambda = 1.8$. For $\rho=10^{-1}$ and $n=100$ the detection threshold appears sooner,
for $\lambda = 1.6$. The experiments illustrate  that $H_1$ is
detected more easily, as the network size $n$ and (density)
parameter $\rho$ increase. Overall the results are 
encouraging with particularly low detection threshold.  For
$\rho=10^{-1}$  and  $n=150$,  Model  $H_1$ is  always  detected  when
present as soon as $\lambda \geq 1.2$.

\paragraph{Computational cost.}
To give some insight into the computational cost of the proposed methodology, we recorded the  running time for the estimation of  $p(H_0|Y)$,  in  various  conditions.   Note  that  the  inference strategy can easily been parallelized. Therefore, to give a fair evaluation, we  applied the  methodology once  for  each network  generated, on  a unique core. The results  presented in Table \ref{table:compcost} were obtained on an Intel Xeon CPU 3.07GHz, for $\lambda=2$ and $\rho=10^{-1}$.  As expected, the  running   time  increases  as   the  network  size   $n$  becomes higher. Similarly, increasing the number  $d$ of covariates induces an additional  computational  effort.  Again,  the  methodology  proposed involves testing  various values  of $K$  (from $1$  to $10$  in these experiments) which can be done in parallel to reduce significantly the running times. If a core is used for  each value of $K$, then the running time is  given essentially  by the  slowest  run, usually  for the  largest value of  $K$. For  information, the  corresponding running  times are also indicated in parenthesis in Table \ref{table:compcost}. 

\begin{table}[ht]
	
		\begin{centering}
			\begin{tabular}{|c|ccc|}
				\hline
				size of the network ($n$) & $d=2$ & $d=5$ & $d=10$ \\ 
				\hline
				100 & 0.47 (0.1) & 0.6 (0.12) & 0.72 (0.14)\\ 
				250 & 3.42  (0.73) & 4.74 (0.88)& 5.97 (1.26)\\ 
				500 & 18.03 (3.73) & 20.28 (4.17) & 24.43 (4.91)\\ 
				\hline
			\end{tabular}
			\caption{Averaged running times (in minutes) for the estimation of $p(H_0|Y)$, for
				various  sizes  $n$  of  networks  and various  values  of  $d$.  In
				parenthesis, the averaged running times (in minutes) for $K=10$.}
			\label{table:compcost}
		\end{centering}
\end{table}

\section{Illustrations}\label{sec:examples}

We applied our approach to analyze a series of networks of various sizes and densities, from social sciences and ecology. For all studies, equal prior probabilities were given for the Models $M_K$ ($K   \geq   2$)   such   that   $p(H_1')=1/2$.   Moreover,   we   set $a_0=b_0=c_0=d_0=e_0=1$. The variational algorithm was run on each network for $K$ between
$1$ and $16$. For each $K$,  the procedure was repeated $20$ times and
the run maximizing the lower bound was selected.

\paragraph{Coding of the covariates.}
The model we propose involves a regression term $x_{ij}^\intercal \beta$ where $x_{ij}$ is a vector of covariates for edge $(i, j)$. In some situations, edge descriptors $x_{ij}$,  such as (phylogenetic, geographic) distances, are actually available. But in many situations, only node descriptors $x_i$ and $x_j$ are available and building an {\sl edge} descriptor $x_{ij}$ from {\sl node} descriptors  is not a straightforward task \citep[see e.g.][]{HGH08}. For all networks (except the blog network to be consistent with \cite{LaR13}), we adopted the following coding rules. Quantitative edge descriptors were treated as quantitative regressors. For quantitative node descriptors, the absolute difference $x_{ij} = |x_i - x_j|$ was used as a quantitative covariate. For ordinal node descriptors $x_i \in \{1, \dots L\}$, we considered the absolute difference $|x_i - x_j|$ but we treated it as a factor, with $L - 1$ levels. Qualitative node descriptors with $L$ levels were transformed into qualitative edge descriptors with $2L$ levels, each node level $\ell$ giving rise to two edge levels: one indicating if both $i$ and $j$ have level $\ell$ and one indicating if either $i$ or $j$ (but not both) has level $\ell$.

\subsection{Description of the datasets}

\begin{paragraph}{Blog network.}
The network is  made of 196 vertices  and was built from  a single day
snapshot  of  political   blogs  extracted   on  14th   October  2006
\citep{articlezanghi2008}. Nodes correspond to blogs and an edge connect
two nodes  if there is  an hyperlink from one  blog to the  other. They
were annotated manually  by the ``Observatoire
Présidentiel'' project such that, for each node, labels are
available. Thus, each node is associated to a political party from the left
wing to  the right  wing and the  status of the  writer is  also given
(political analyst or not).  This data set has been studied in a series of works
\citep{articlezanghi2008,articlelatouche2011,latouche2014} where
all the  authors pointed out the  crucial role of the  labels in the
construction of the network. We considered a set of three covariates  $x_{ij}=(x_{ij}^1,x_{ij}^{2},x_{ij}^{3})\in
\mathbb{R}^{3}$          artificially
constructed to analyze the influence of both
the political parties and the writer status. We set $x_{ij}^{1}=1$ if blogs $i$ and
$j$ have the same labels, 0 otherwise. Moreover, $x_{ij}^{2}=1$ if one of the two blogs $i$ and $j$ is written by
political analysts, $0$ otherwise. Finally, $x_{ij}^{3}=1$ if both are
written by political analysts, $0$ otherwise. 
\end{paragraph}

\begin{paragraph}{Tree network.}
  This data set was first introduced by \cite{VPD08} and further studied
in \cite{articlemariadassou2010}. We considered the tree network which
describes the interactions between $51$ trees where two trees interact
if  they   share  at   least  one   common  fungal   parasite.   Three
{quantitative edge descriptors}
are available  characterizing the  genetic, geographic,  and taxonomic
distances between the tree species.
\end{paragraph}

\begin{paragraph}{Karate network.}
The karate data set describes the friendships between a subset of 34 members of a
karate club at a university in the  US, observed from 1970 to 1972. It
was originally  studied by \cite{zachary1977}.  When the study
started, an incident occurred between  the club president and a karate
instructor, over the price of the karate lessons. The entire club then
became divided over this issue, as time passed. The network is made of
four known groups characterized by a node
  qualitative descriptor,  taking four possible values,  for each node
  in the network. 
\end{paragraph}

\begin{paragraph}{Florentine marriage network.}

We considered  the data  set analyzed  by \cite{breiger1986}  in their
study of local role analysis  in social networks. It characterizes the
social relations among 16 Renaissance Florentine families and was built
by John Padgett from historical documents. Two nodes are linked is the
two families share marriage alliances. Three {quantitative node} covariates are provided
for each family, namely the family's  net wealth in 1472 in thousands of
lira, the family's  number of seats on the civic  councils held between
1282 and 1344, and the family's  total number of business and marriage
ties in the entire data set. 
  \end{paragraph}

\begin{paragraph}{Florentine business network.}
 This data set is similar to the Florentine marriage network described
 previously except that edges
 now describe business ties between families.  We considered exactly the same
 covariates. 
\end{paragraph}

\begin{paragraph}{Faux Dixon High network.}
Contrary to  all networks  presented in  this work,  this data  set is
directed and  therefore we  employed the  inference algorithm  for the
directed case, as presented in the supplementary materials.  
This  network  characterizes  the (directed)  friendship  between  248
students.  It results  from  a simulation  based  upon an  exponential
random graph model fit \citep{handcock2008} to data from one school community from the AddHealth Study, Wave I (Resnick et al., 1997).  Node
covariates  are provided,  namely the  grade,  sex, and  race of  each
student.  The grade {ordinal} attribute has values 7 to 12, indicating each
student's grade in school. Moreover, the race {qualitative} attributes can take {4} values. 
\end{paragraph}

\begin{paragraph}{CKM.}
This data set was created  by \cite{Burt1987} from the data originally
collected   by   \cite{Coleman1966}.   The   network   we   considered
characterizes  the  friendship  relationships among  physicians,  each
physician being asked to name  three friends. The physicians were also
asked to answer to a series of questions regarding their profession. We focused
here on  13 questions corresponding  to node covariates among  which four
are qualitative descriptors: city of practices (4 values), discussion
with other  doctors (3 values), speciality  in a field of  medicine (4
values),  proximity  with  other  physicians  (4  values).  All  other
node covariates were treated as quantitative variables.  Note that we imputed
the missing values in the data set using the missMDA R package \citep{husson2016}.
\end{paragraph}

\begin{paragraph}{AddHealth 67.}
This data  set is  related to the  Faux Dixon  network described
  previously. However,  it was constructed  from the original  data of
  the  AddHealth  study,  and  not simulated  from  any  random  graph
  model.   The  AddHealth   study   was   conducted  using   in-school
  questionnaires, from 1994  to 1995. Students were  asked to designate
  their friends and  to answer to a series of  questions. Results were
  collected in schools from 84 communities. In our study, we considered a network
  associated to school community 67 which characterizes the undirected
  friendship relationships between 530 students. As for the Faux Dixon
  network, three  node covariates  are available. The  sex qualitative
  covariate takes  two values.  Moreover, the grade  ordinal attribute
  has  values from  7  to  12. However,  contrary  to  the Faux  Dixon
  network,  five  values  are  present   in  the  data  for  the  race
  qualitative attribute.

\end{paragraph}

\subsection{Results}

\begin{table}[!]
\begin{centering}
\begin{tabular}{|l|cccc|}
  \hline
 Network & size ($n$) & nb. covariates ($d$) & density &$\hat{p}(H_0|Y)$ \\ 
  \hline
Blog & 196 & 3 &0.075&  3.60e-172 \\
Tree & 51 & 3 & 0.54 & 2.36e-115\\
Karate & 34 & 8 &0.14 & 3.38e-2 \\
Florentine (marriage) & 16 & 3 & 0.17 & 0.995 \\
Florentine (business) & 16 & 3 & 0.125 & 0.991 \\
Faux Dixon High & 248 & 17 & 0.02 & 1 \\
CKM & 219 & 39 & 0.015 & 1 \\
AddHealth 67 & 530 & 21 & 0.007& 2.10e-25 \\
     \hline
\end{tabular}
\caption{Estimation of $p(H_0|Y)$, for
  the eight networks considered.} 
\label{table:realnet_res}
\end{centering}
\end{table}

\begin{figure}[!]
 \centering
 \caption{\label{fig:reject} 
 Estimation of the blog (top) and tree (bottom) networks residual structure without (left) and with (right) covariates.}
 \begin{tabular}{cc}
  & \\
  \includegraphics[width=.45\textwidth]{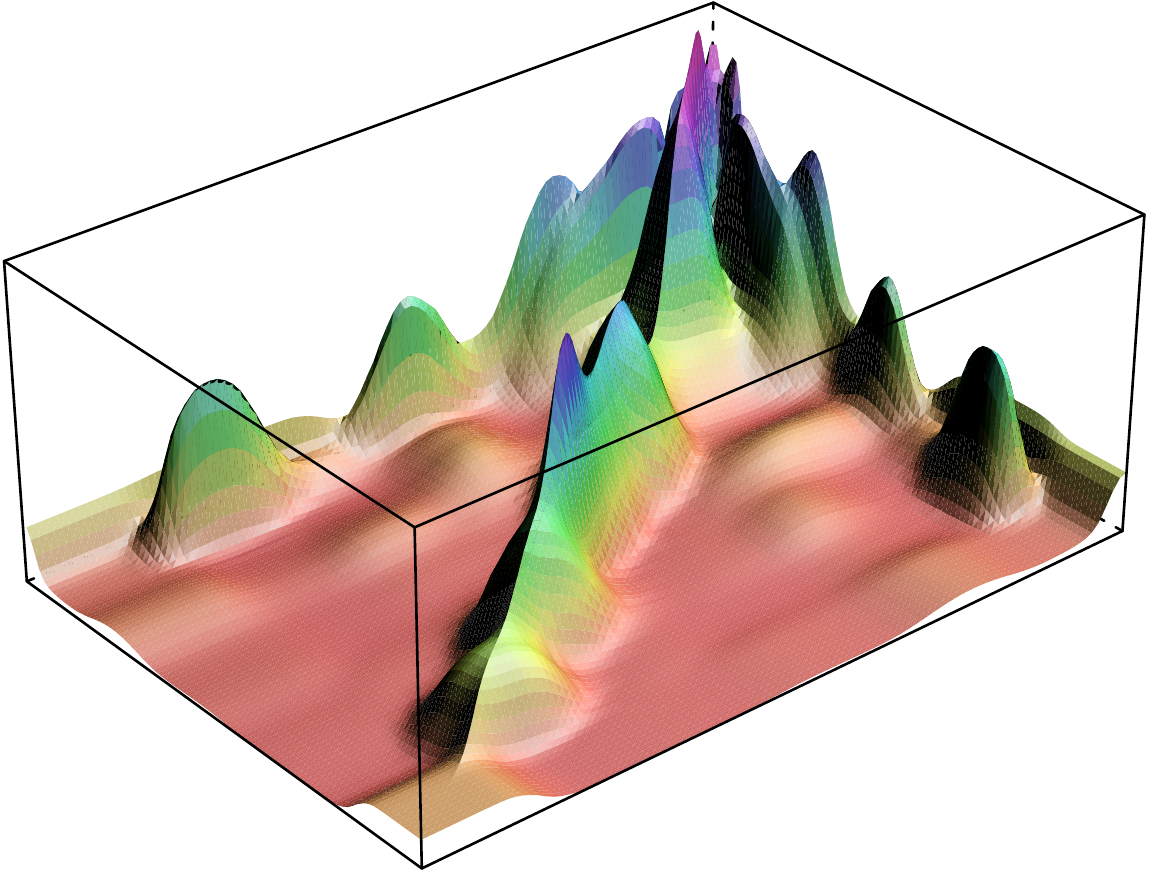} &
  \includegraphics[width=.45\textwidth]{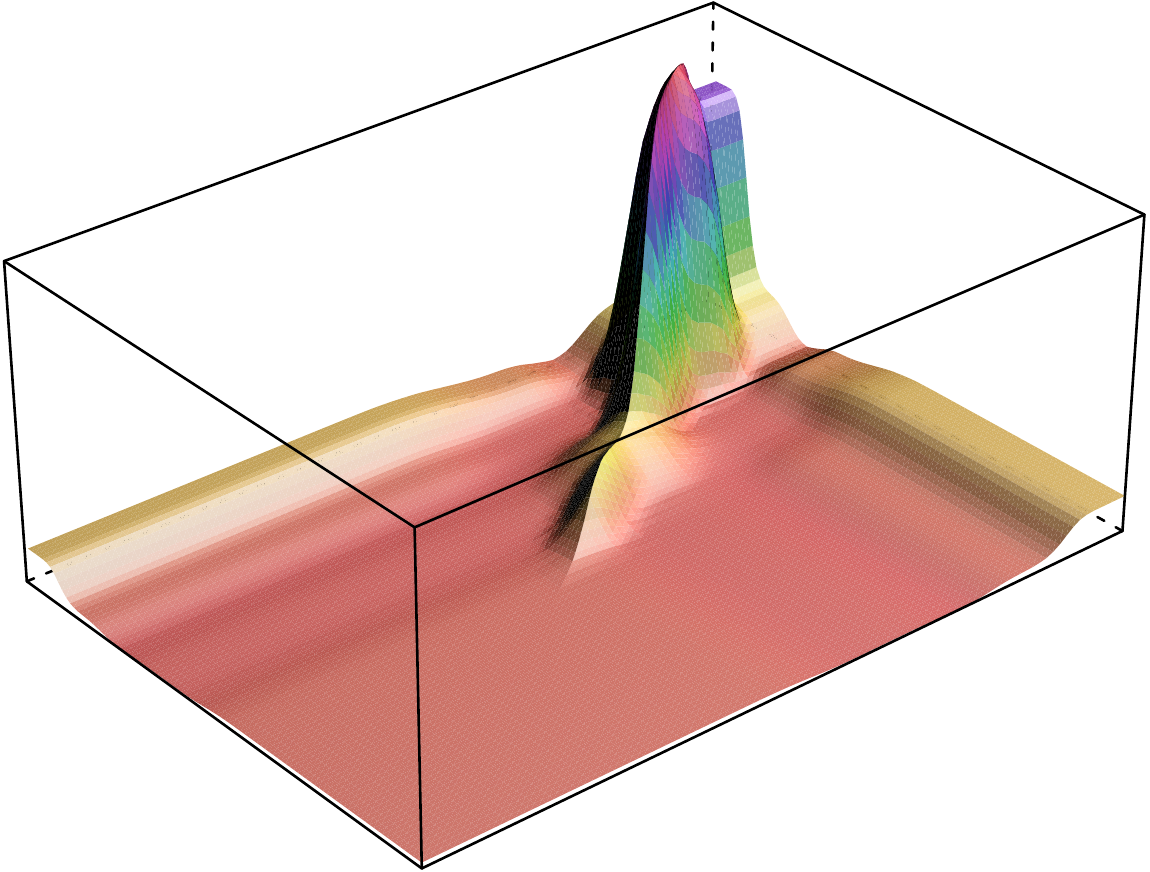} \\
  \includegraphics[width=.45\textwidth]{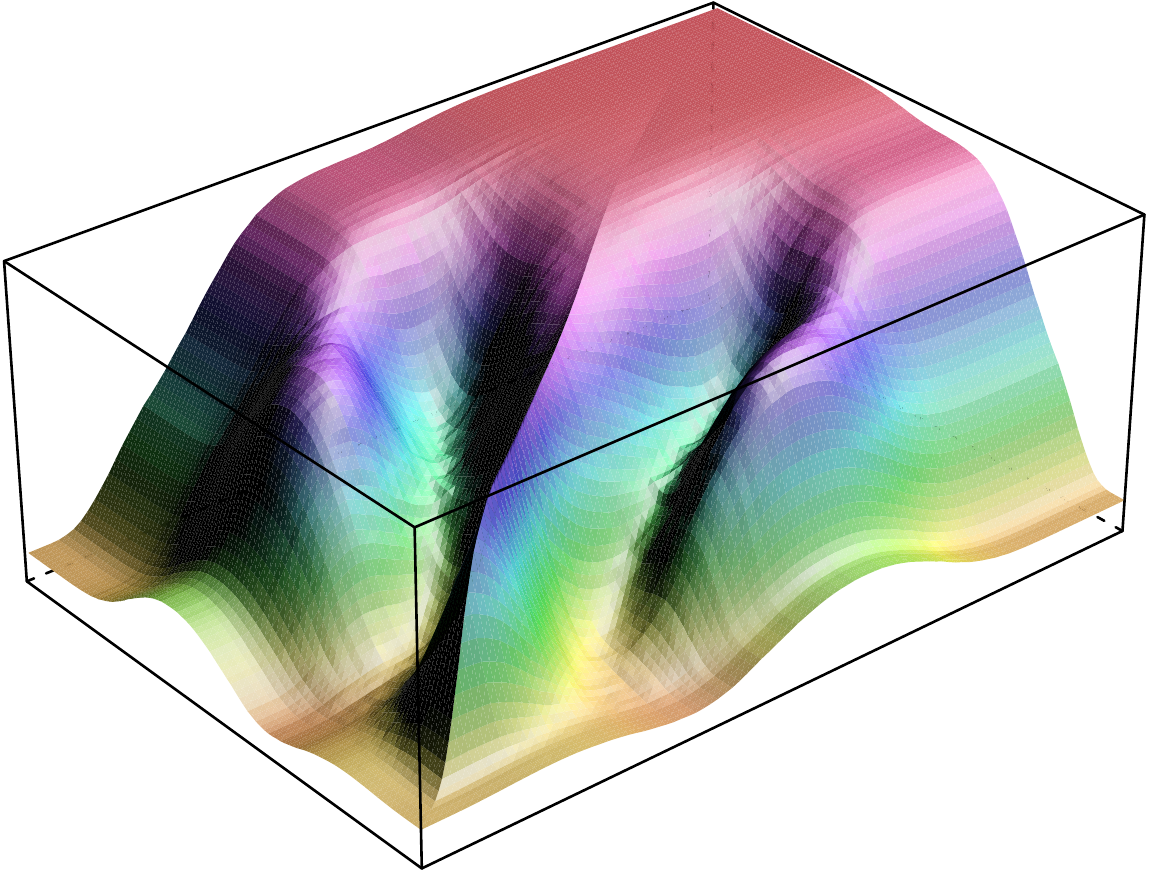} &
  \includegraphics[width=.45\textwidth]{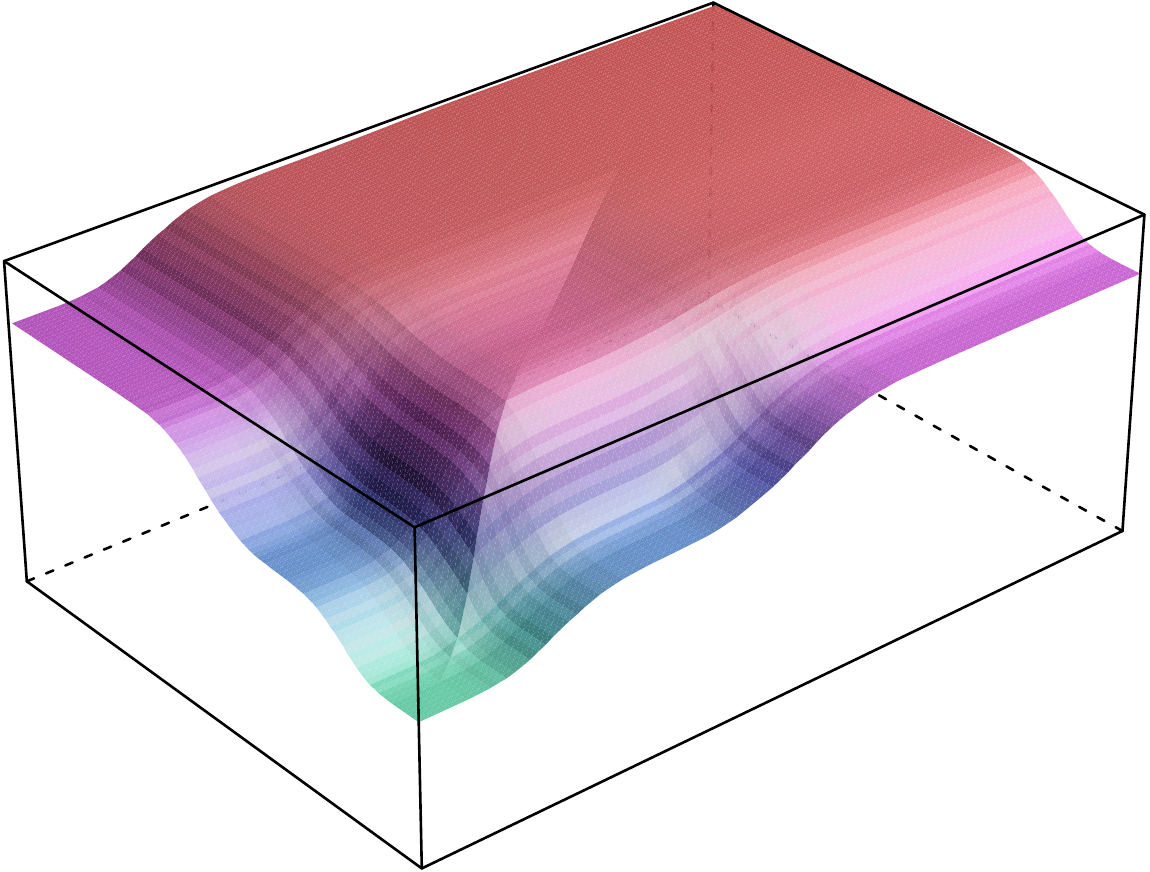} 
 \end{tabular}
\end{figure}

\begin{figure}[!]
 \centering
 \caption{\label{fig:reject2} 
 Estimation of the karate (top) and AddHealth (bottom) networks residual structure without (left) and with (right) covariates.}
 \begin{tabular}{cc}
  & \\
  \includegraphics[width=.45\textwidth]{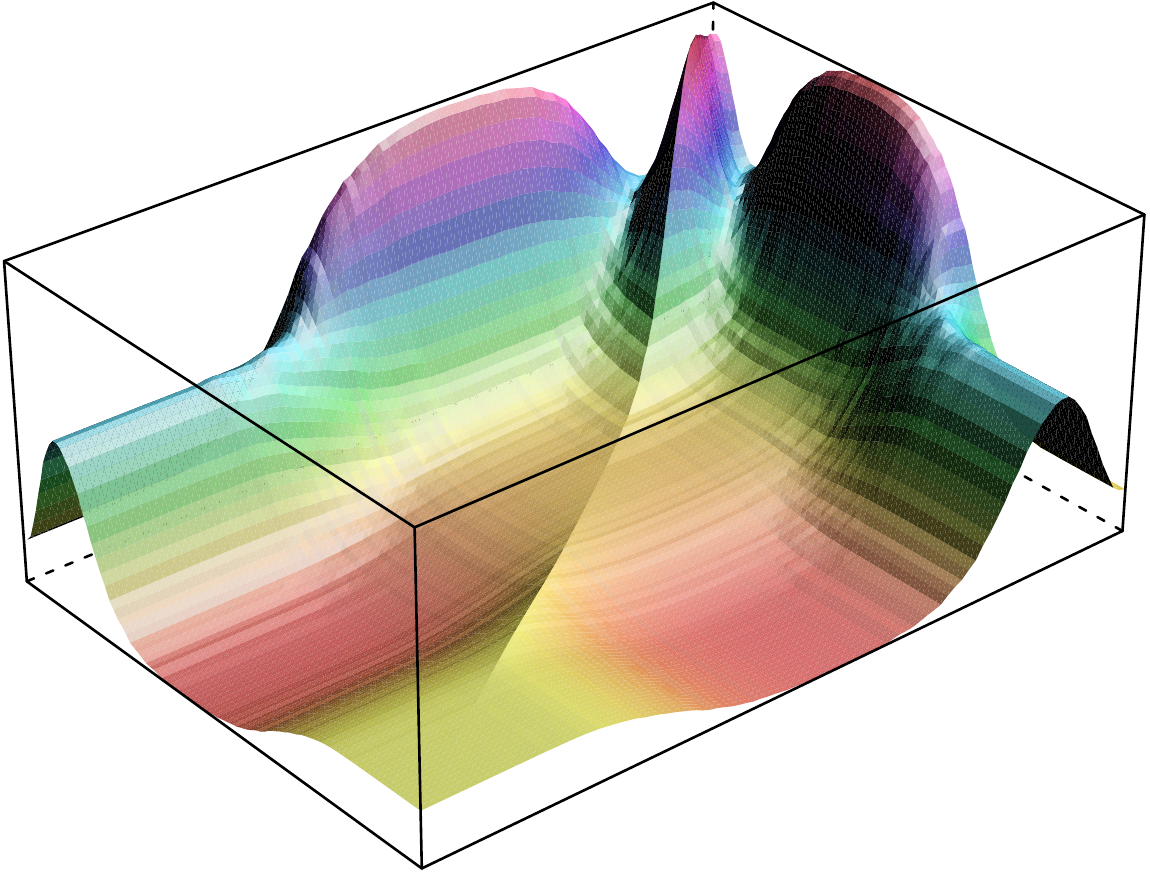} &
  \includegraphics[width=.45\textwidth]{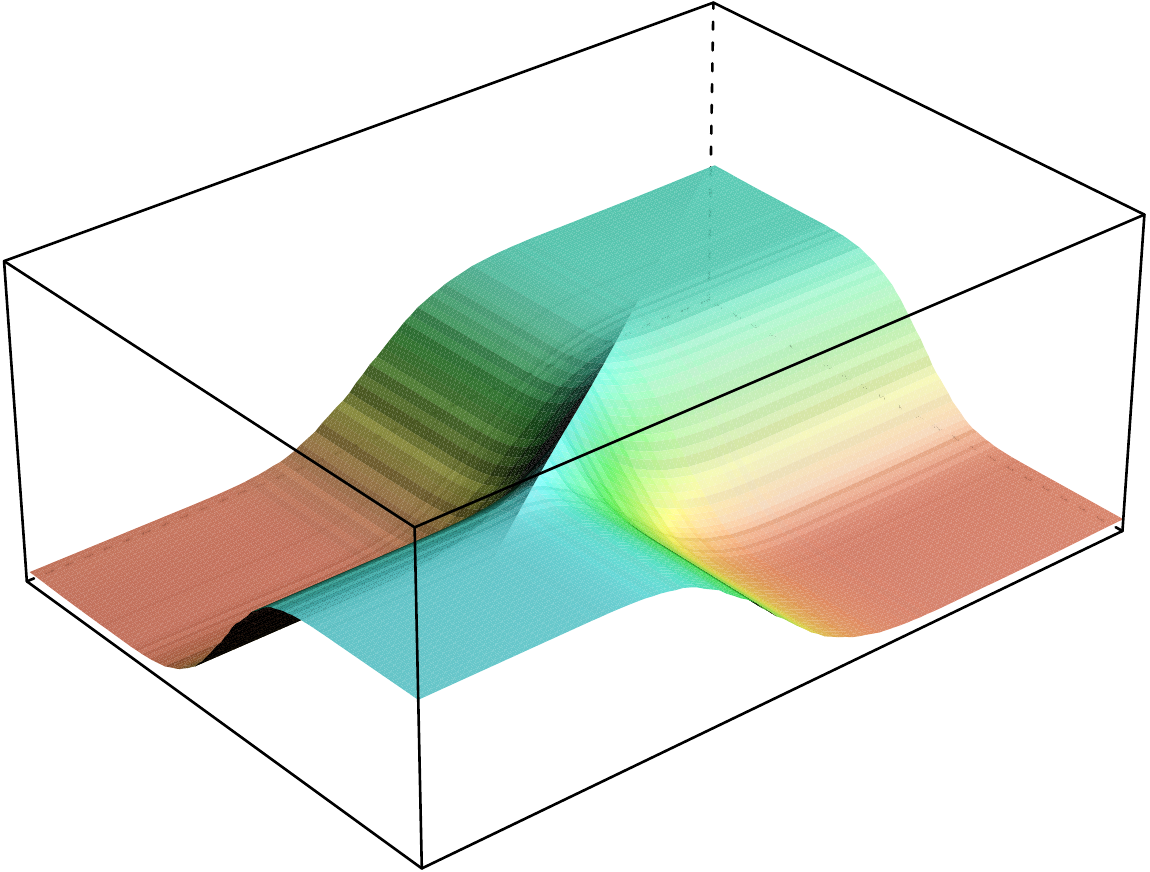} \\
  \includegraphics[width=.45\textwidth]{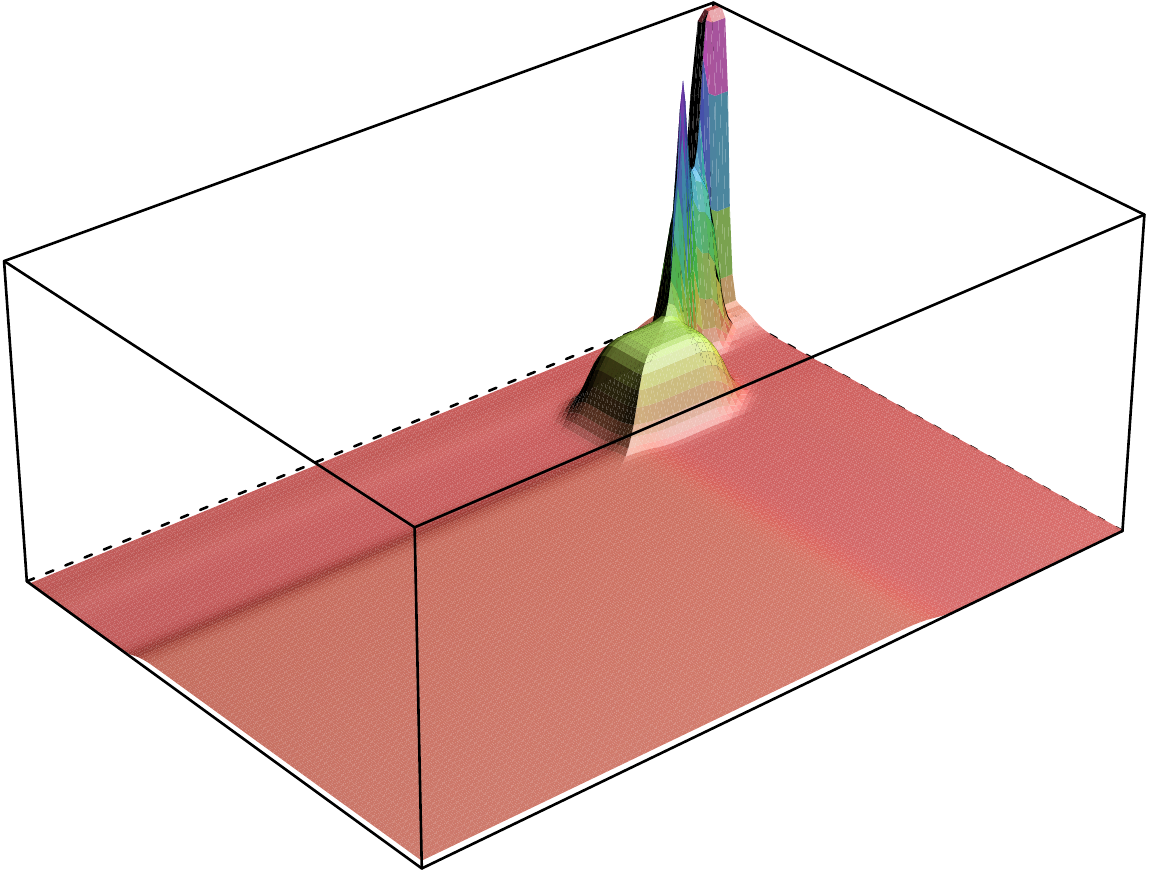} &
  \includegraphics[width=.45\textwidth]{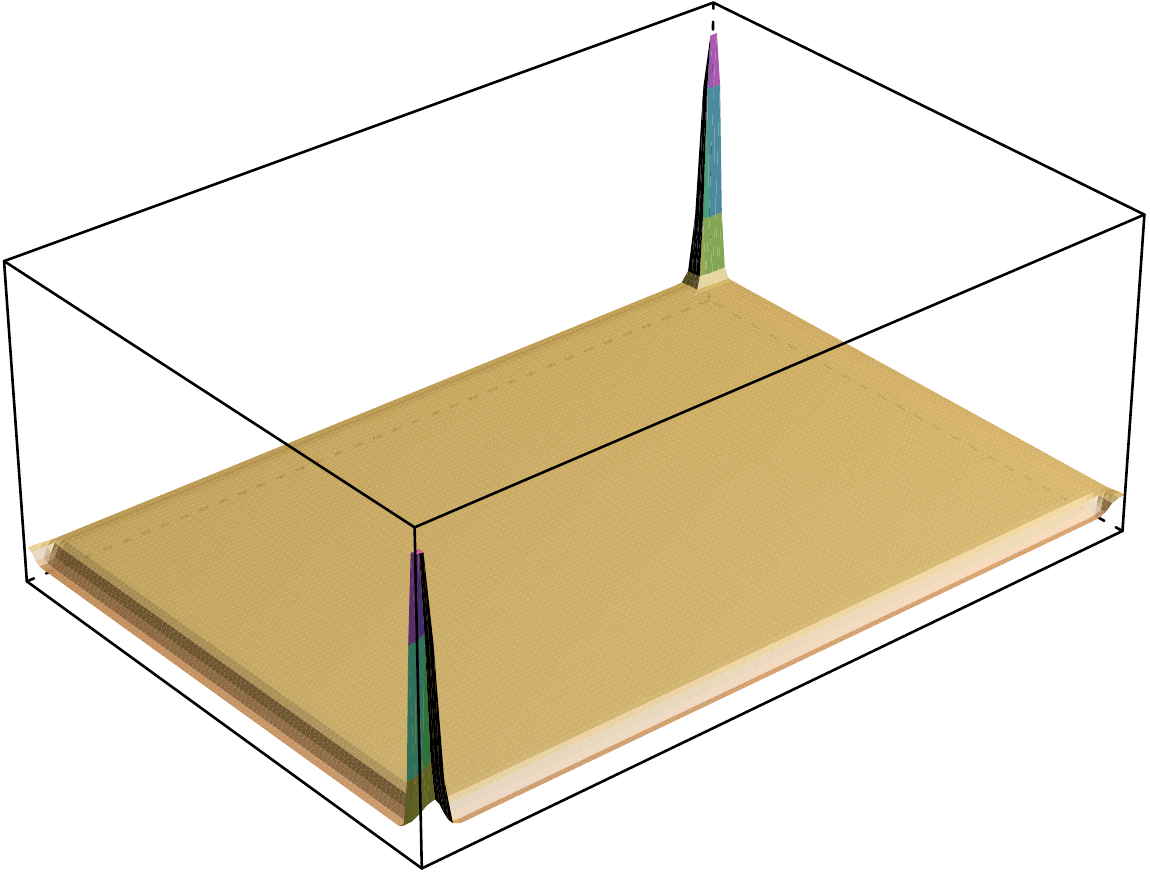} 
 \end{tabular}
\end{figure}

\begin{figure}[!]
 \centering
 \caption{\label{fig:accept} 
 Estimation  of the  Florentine  marriage  (top), Florentine  business
 (middle), and CKM networks residual structure without (left) and with (right) covariates.}
 \begin{tabular}{cc}
  & \\
  \includegraphics[width=.45\textwidth]{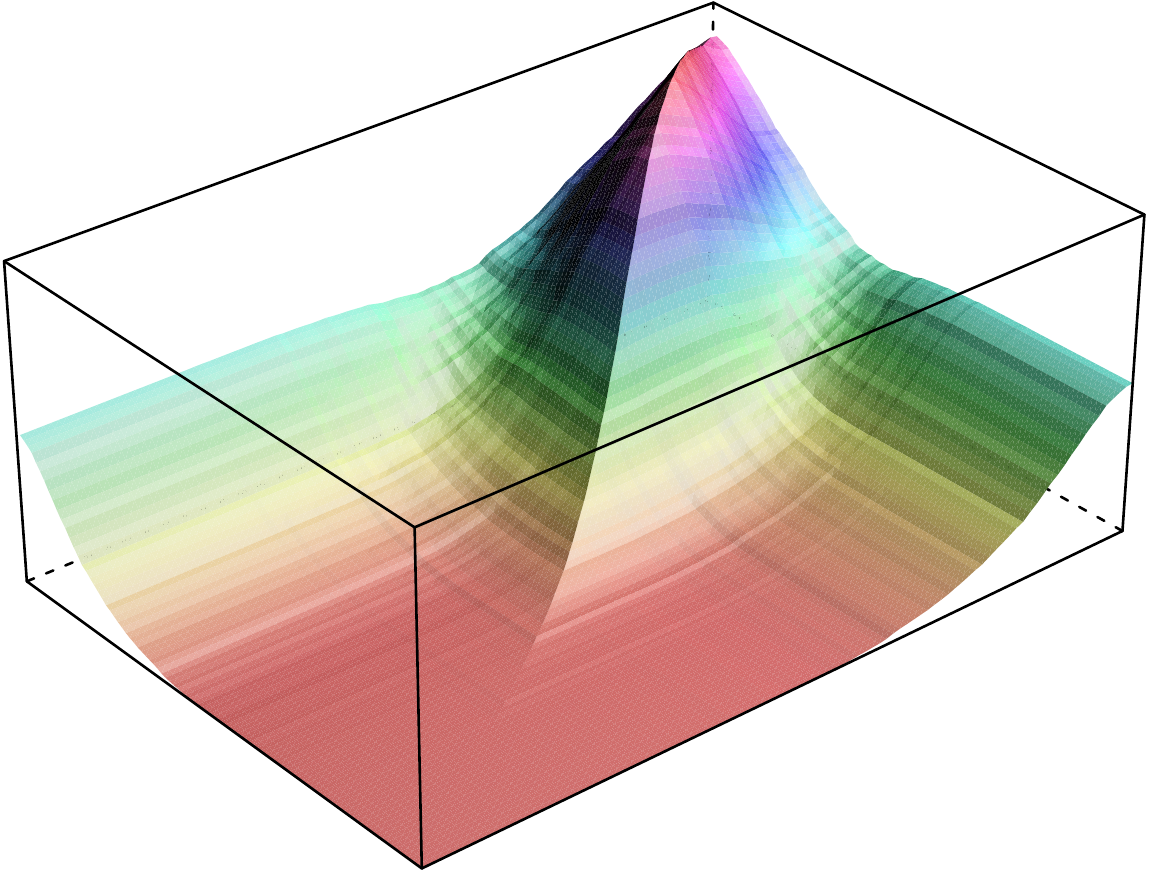} &
  \includegraphics[width=.45\textwidth]{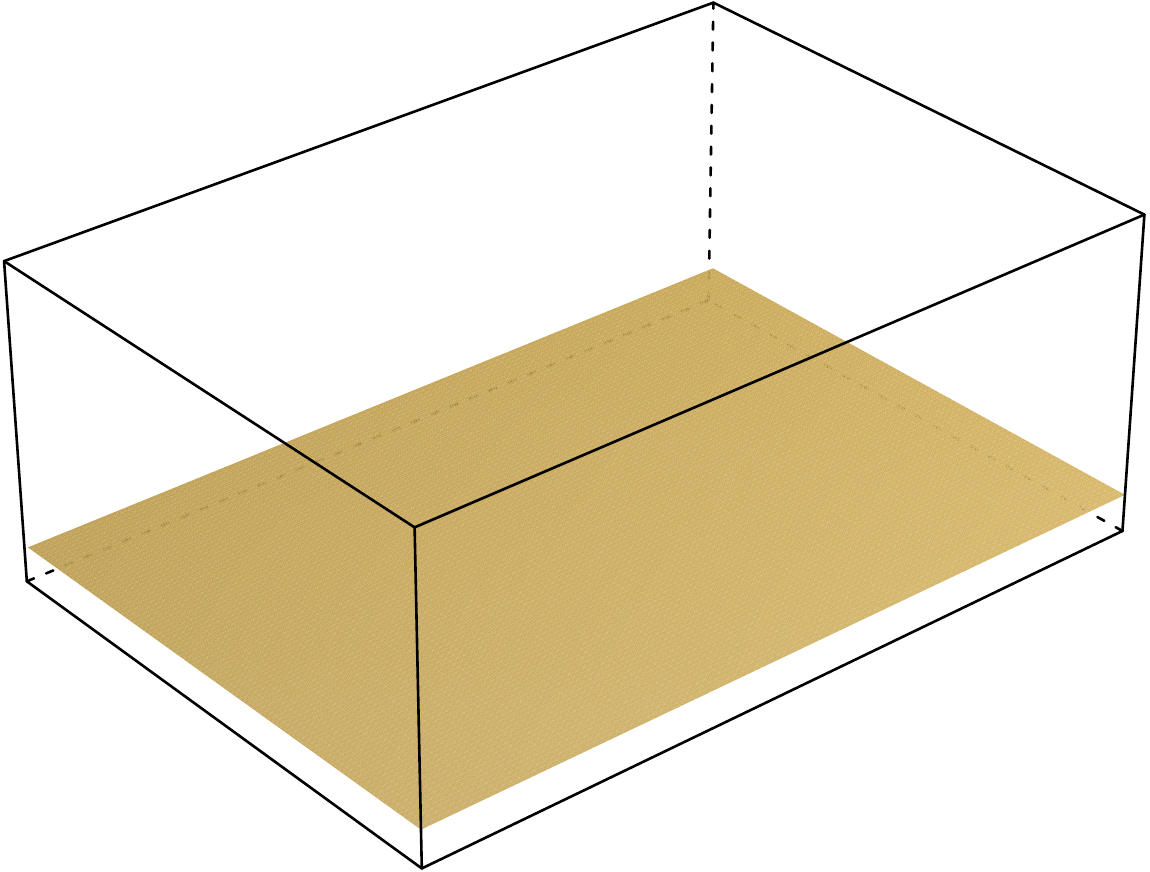} \\
  \includegraphics[width=.45\textwidth]{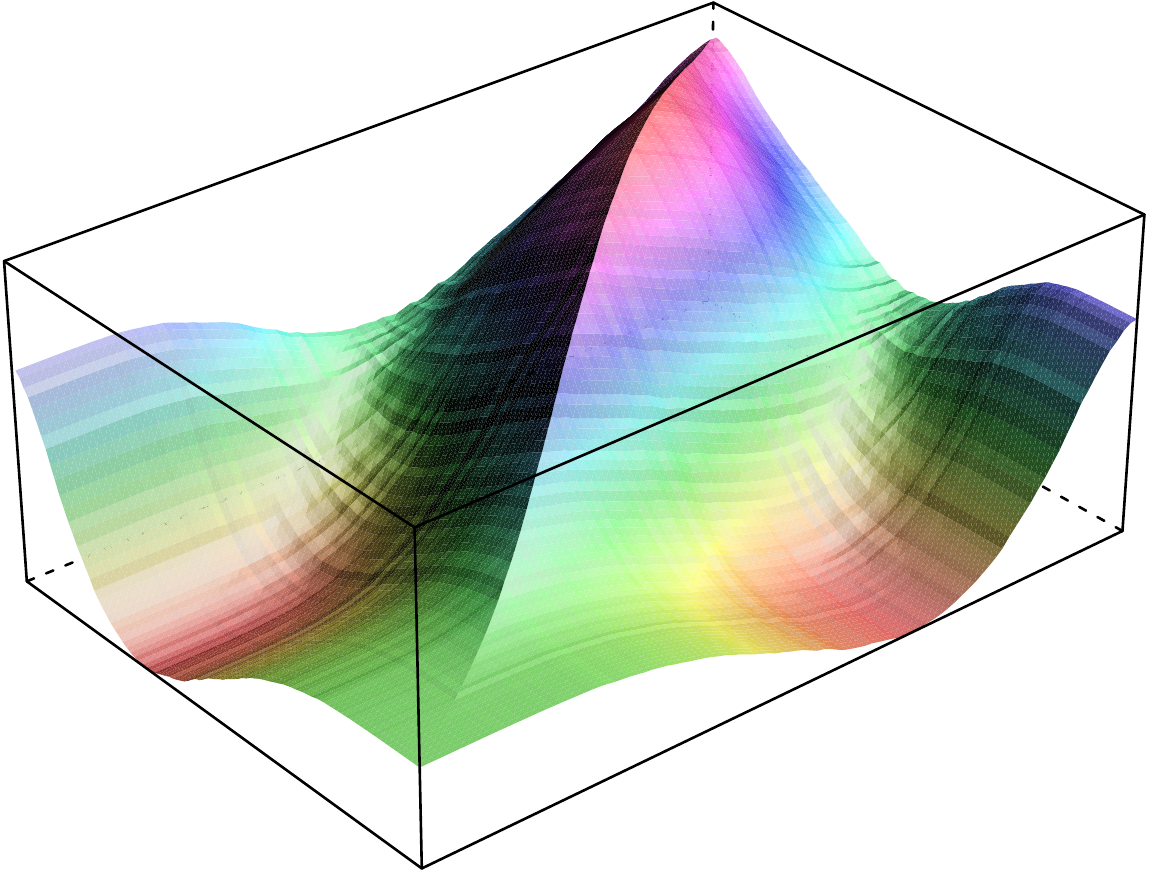} &
  \includegraphics[width=.45\textwidth]{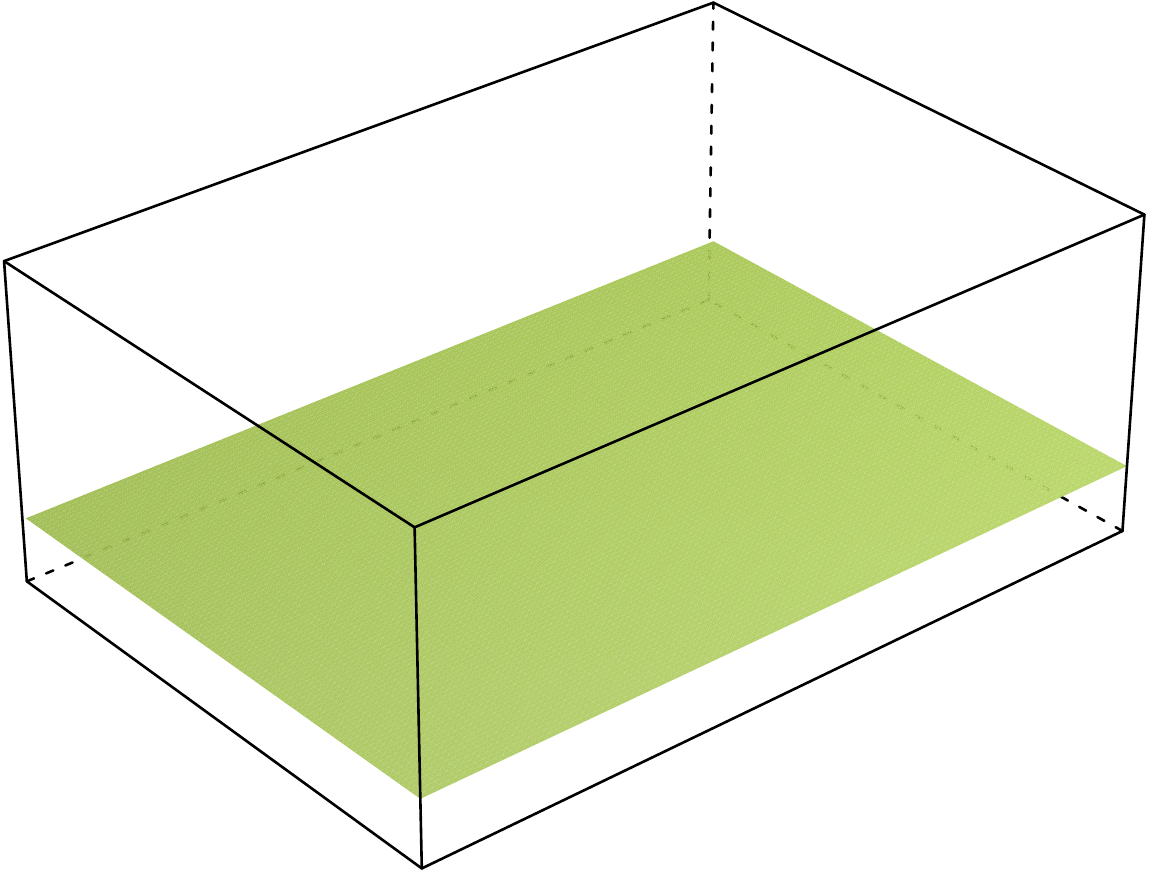} \\
 \includegraphics[width=.45\textwidth]{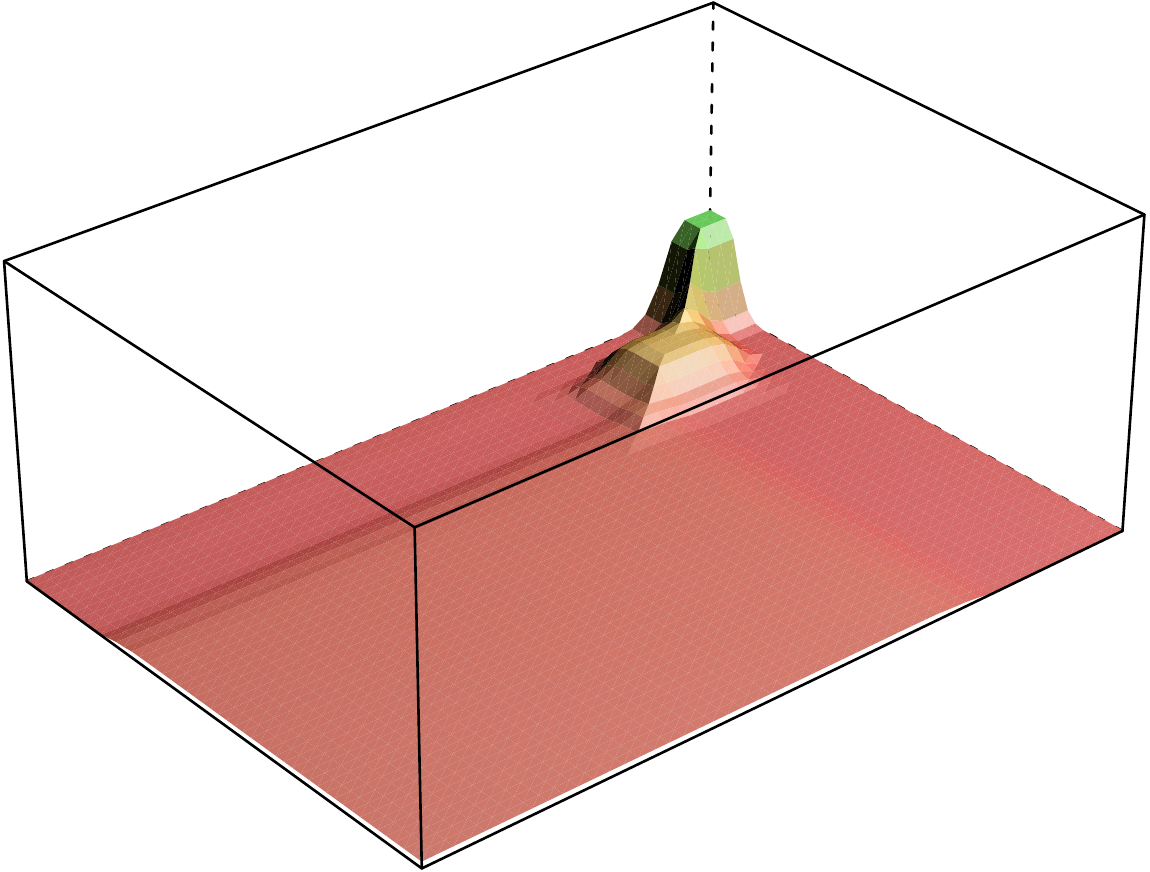} &
  \includegraphics[width=.45\textwidth]{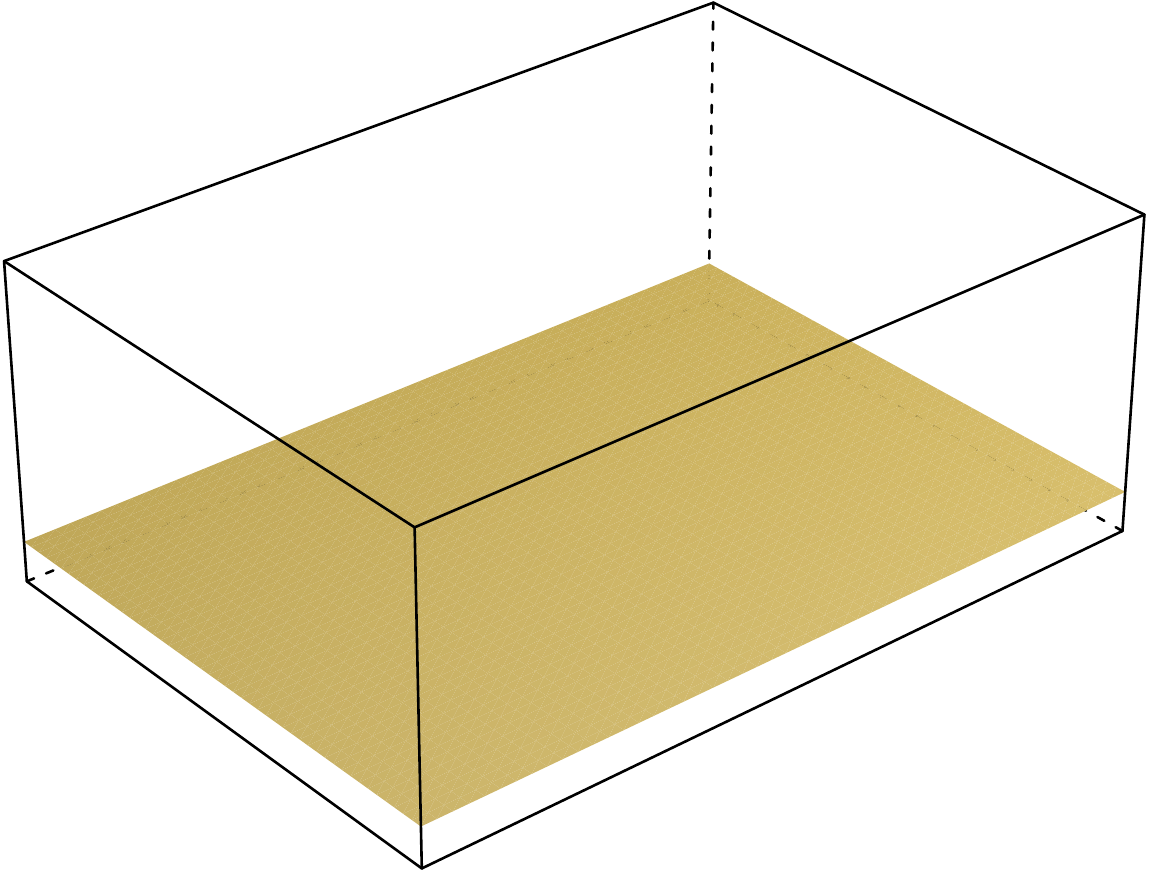} \\
 \end{tabular}
\end{figure}

The estimated values of $p(H_0|Y)$ for all networks are presented in Table
\ref{table:realnet_res}. For illustration  purposes, the estimations of
the  residual structures  $g \circ  \hat{\phi}$ are  also provided  in
Figures {\ref{fig:reject}, \ref{fig:reject2}, and \ref{fig:accept}}.  In practice, we used Proposition \ref{prop:ma1} to estimate
$\hat{\phi}$ and then applied $g(\cdot)$ to obtain  graphon-like surfaces. There  is   no  standard  definition  of
$W$-graph  models in  the directed  case and  therefore, for  the Faux
dixon high network, only
the estimation of $p(H_0|Y)$ is given.

As shown  in Table  \ref{table:realnet_res}, Model $H_0$  was rejected
for the blog, tree, karate and AddHealth networks. Indeed,
we obtained values of $\hat{p}(H_0|Y)$ close to zero for the {four} data sets,
indicating that the {corresponding} covariates cannot explain entirely the construction of
these  networks.  For  the  blog  network, we  can  observe in  Figure
\ref{fig:reject} (top right) that $g \circ
\hat{\phi}$ is not  constant which is coherent with  Model $H_0$ being
rejected. We also give in this figure (top left) the estimated residual structure
without taking the covariates into account ($d=0$). Clearly, the shape
of $g \circ \hat{\phi}$ is simpler when $d=3$.  In  particular, many of the  hills on the diagonal  vanish when
adding  the covariates.  Thus, the  covariates help  in studying  and
explaining parts of the network. However, they are not sufficient and some  of the 
heterogeneity observed in the network cannot be explained by political
parties and writer  status. Similar conclusions can be  drawn from the
tree , karate, and AddHealth networks (Figure
\ref{fig:reject} and Figure
  \ref{fig:reject2}). Indeed, the {terms} $g \circ \hat{\phi}$ {simplify} when adding the
covariates but remain non constant.  {In particular, for the tree
  network considered, this means that the} interactions between
trees through common fungal parasite cannot be entirely explained by the
distances available  which is consistent with a these from
\cite{articlemariadassou2010} who describe a residual heterogeneity in
the valued version  of this network, after taking  the covariates into
account. 

For all other networks considered, model $H_0$ was chosen.  Indeed, for the
Florentine marriage and business networks, we found
$\hat{p}(H_0|Y)=0.995$            and           $\hat{p}(H_0|Y)=0.991$
respectively.  As expected, the residual structures $g \circ
\hat{\phi}$ were found constant when adding the covariates (Figure \ref{fig:accept}). {Moreover}, the variational approach led to
$\hat{p}(H_0|Y)=1$, for the Faux Dixon High {and CKM networks}. Thus, the statistical framework we propose shows that no other effect than these of the covariates contributes significantly to explain the structure of these networks. In other words, once corrected for the covariates, no residual heterogeneity is observed among the interactions. 

\section{Conclusion}

In this paper we proposed a framework to assess the goodness of fit of
logistic models for binary networks. Thus, we added a generic
term, related  to the  graphon function of  $W$-graph models,  to the
logistic regression model. The corresponding new model was approximated
with a series of models  with blockwise constant residual structure. A
Bayesian  procedure  was  then considered  to  derive  goodness-of-fit
criteria. All these  criteria depend on marginal  likelihood terms for
which   we    did   provide    estimates   relying    on   variational
approximations.  The  first  approximation  was  obtained  using  a  variational
decomposition   while  the   second  involves   a  series   of  Taylor
expansions. The approach was tested on toy data sets and
encouraging results were obtained. Finally, it was used to analyze {eight}
 networks  from  social  sciences and  ecology.  We  believe  the
methodology  has  a  large  spectrum of  applications  since  covariates are often given when analyzing binary networks.

\bibliographystyle{chicago}

\newpage
\spacingset{1.45} 
\begin{center}
{\large\bf SUPPLEMENTARY MATERIAL}
\end{center}

\begin{description}

\item[Appendix:] Give all proofs of the paper. (Appendix.pdf)

\item[Directed  case:]  Describe  the   inference  procedure  for  the
  directed case. (Directed.pdf)

\end{description}


\end{document}